\documentclass[pdflatex,sn-mathphys-ay,iicol]{sn-jnl}

\usepackage{graphicx}%
\usepackage{multirow}%
\usepackage{amsmath,amssymb,amsfonts}%
\usepackage{amsthm}%
\usepackage{mathrsfs}%
\usepackage[title]{appendix}%
\usepackage{xcolor}%
\usepackage{textcomp}%
\usepackage{manyfoot}%
\usepackage{booktabs}%
\usepackage{algorithm}%
\usepackage{algorithmicx}%
\usepackage{algpseudocode}%
\usepackage{listings}%
\usepackage{tikz}
\usetikzlibrary{arrows.meta, positioning, calc}
\usepackage{subcaption}
\usepackage{ulem}
\usepackage{hyperref}

\newcommand{\g}{\,|\,}
\newcommand{\bseta}{\boldsymbol{\eta}}
\newcommand{\btheta}{\boldsymbol{\theta}}
\newcommand{\by}{\boldsymbol{y}}
\newcommand{\bphi}{\boldsymbol{\phi}}
\newcommand{\bc}{\boldsymbol{c}}
\newcommand{\bci}{\bc^{(i)}}
\newcommand{\bmu}{\boldsymbol{\mu}}
\newcommand{\btau}{\boldsymbol{\tau}}
\newcommand{\bxi}{\boldsymbol{\xi}}
\newcommand{\bkappa}{\boldsymbol{\kappa}}
\newcommand{\tbphi}{\tilde{\boldsymbol{\phi}}}

\newcommand{\ind}{^{(i)}}

\theoremstyle{thmstyleone}%
\theoremstyle{thmstyletwo}%

\theoremstyle{thmstylethree}%

\begin{document}

\title[SBI for NLMEMs]{Simulation-based inference for stochastic nonlinear mixed-effects models with applications in systems biology}

\author[1]{\fnm{Henrik} \sur{Häggström}}\email{henhagg@chalmers.se}

\author[1]{\fnm{Sebastian} \sur{Persson}}\email{sebastian.persson@crick.ac.uk}

\author[1]{\fnm{Marija} \sur{Cvijovic}}\email{marija.cvijovic@chalmers.se}

\author*[1]{\fnm{Umberto} \sur{Picchini}}\email{picchini@chalmers.se}

\affil[1]{\orgdiv{Department of Mathematical Sciences}, \orgname{Chalmers University of Technology and the University of Gothenburg}, \city{Gothenburg}, \country{Sweden}}

\abstract{The analysis of data from multiple experiments, such as observations of several individuals, is commonly approached using mixed-effects models, which account for variation between individuals through hierarchical representations. This makes mixed-effects models widely applied in fields such as biology, pharmacokinetics, and sociology. In this work, we propose a novel methodology for scalable Bayesian inference in hierarchical mixed-effects models.  Our framework first constructs amortized approximations of the likelihood and the posterior distribution, which are then rapidly refined for each individual dataset, to ultimately approximate the parameters posterior across many individuals. The framework is easily trainable, as it uses mixtures of experts but without neural networks, leading to parsimonious yet expressive surrogate models of the likelihood and the posterior. We demonstrate the effectiveness of our methodology using challenging stochastic models, such as mixed-effects stochastic differential equations emerging in systems biology-driven problems. However, the approach is broadly applicable and can accommodate both stochastic and deterministic models. We show that our approach can seamlessly handle inference for many parameters. Additionally, we applied our method to a real-data case study of mRNA transfection. When compared to exact pseudomarginal Bayesian inference, our approach proved to be both fast and competitive in terms of statistical accuracy.}


\maketitle

\begingroup
\small
\noindent\textbf{Accepted manuscript notice.}
This version of the article has been accepted for publication, after peer review (when applicable) but is
not the Version of Record and does not reflect post-acceptance improvements, or any corrections.
The Version of Record is available online at:
\url{http://dx.doi.org/10.1007/s11222-026-10850-8}.
\par
\endgroup

\section{Introduction}\label{sec:intro}
The analysis of data arising from multiple experiments — such as observations of individuals across various domains (e.g., humans, animals, cells, or trees) — has traditionally been approached using mixed-effects models \citep{diggle2002analysis,lavielle2014mixed}. These models account for individual variability by treating model parameters as random variables that vary between subjects, allowing a hierarchical representation of the data. This structure effectively disentangles different sources of variability, distinguishing intra-individual fluctuations from between-individual differences. Mixed-effects modeling is widely applied across diverse fields, including biology, pharmacokinetics and pharmacodynamics, forestry, sociology, and many other disciplines where understanding variability across individuals is essential \citep{davidian2003nonlinear}. The literature to tackle the inference problem for mixed-effects models is vast. However, options reduce substantially when stochastic dynamical models are considered due to considerably increased theoretical and computational difficulties.
In this work, we present a new strategy for Bayesian inference in hierarchical nonlinear stochastic models with mixed-effects. Our approach  provides fast to train \textit{semi-amortized} approximations to both the likelihood function and the posterior distribution. We show that this makes our methodology scalable for an increasing number of individuals. In particular, we consider stochastic models with time-dynamics, with a focus on stochastic differential equations (SDEs) with mixed-effects. However, \textcolor{black}{we consider SDE models}  merely to provide illustrative case studies, as the \textcolor{black}{statistical inference} methodology we propose is agnostic to the specific type of model used to describe data,  \textcolor{black}{the only requirement being the availability of a generative model to simulate data}. 

Inference for stochastic modelling with mixed-effects is a challenging area due to the difficulty in (numerically) integrating out the latent quantities that enter the likelihood function. For example, models for stochastic chemical reactions that are common in systems biology involve either exact simulators (e.g. the Stochastic Simulation Algorithm, \citealp{gillespie1977exact}, the Extrande method, \citealp{voliotis2016stochastic}) or approximate simulators (\citealp{gillespie2000chemical}, \citealp{gillespie2007stochastic}), and all these make the likelihood function \textit{intractable}, that is, it is not possible to evaluate the likelihood exactly and is computationally hard to approximate. For example, when the model is an SDE, and observations are available at discrete time points, the unavailability of closed-form transition densities makes the likelihood function intractable (except for the simplest toy models). The statistical literature to tackle such difficulty is vast, see eg. \cite{craigmile2023statistical} for a recent review. On top of such difficulty, when SDEs are embedded in a mixed-effects framework, the problem becomes even harder due to the increased dimension of the integration problem. Although the literature for mixed-effect SDEs is available, as collected in \cite{sdemem-webpage}, this is very specialized, as methods either tackle very specific models, lacking generality and well-maintained software, or are general but computationally very intensive, e.g. when based on particle-filters \citep{botha2021particle, wiqvist2021, persson2022}.

In this work, we provide a general approach for Bayesian inference in mixed-effects models having intractable likelihoods, using simulation-based inference (SBI) (see \citealp{cranmer2020frontier} for a review). 
The appeal of SBI methods is
that these only require forward-simulation of the model, rather than the evaluation of a
potentially complicated expression for the likelihood function, assuming it is available.
This allows for approximate frequentist and Bayesian inference, whenever running the model simulator at many values of a parameter $\btheta$ is computationally not too onerous. Indeed, in SBI, simulated data $\by$ are generated as $\btheta\rightarrow \mathcal{M}(\btheta)\rightarrow \by$, where the \textit{simulator} $\mathcal{M}$ can be any generative model. 
Provided with many pairs of simulated $\btheta$'s and $\by$'s, it is possible to build inference for given observed data $\by_o$, in absence of a readily available expression for the likelihood function $p(\by_o|\btheta)$, as we briefly summarize in Section \ref{sec:background}. 
In recent years many SBI methods have focused on using deep neural networks to approximate conditional densities (``neural conditional density estimation''), for example to provide approximations to the posterior $p(\btheta|\by_o)$, the likelihood $p(\by_o|\btheta)$, or both, see Section \ref{sec:background} for key references. In our work, we obtain scalable and accurate inference for complex stochastic models with mixed-effects,  without using neural conditional density estimation (NCDE). \textcolor{black}{The relevance of using probabilistic frameworks that are more parsimonious than neural networks can result in easier analytic tractability: in our case, we use Gaussian mixture models (GMMs) to approximate conditional densities, and the decades-long research on GMMs \citep{fruhwirth2019handbook} provides estimators via ad-hoc algorithms for their fitting, such as expectation-maximization \citep{dempster1977maximum}. Moreover, the estimators obtained from a specific conditional density (say the likelihood function), when expressed via a GMM,  can easily be transferred to other conditional densities (eg the posterior) using closed-form algebraic operations.}

We build on the \textit{SEquential Mixture Posterior and Likelihood Estimation} (SeMPLE) methodology \citep{haggstrom2024}, which efficiently fits training data using a GMM via an expectation-maximization algorithm. The fitted Gaussian mixture provides, simultaneously, a closed-form deterministic approximation to both the likelihood and the posterior, which can both be evaluated and sampled from in a Gibbs sampler. For the case where $M$ individuals are considered, with corresponding observed data $\by^{(i)}_o$ ($i=1,...,M$), the surrogates of the likelihood and the posterior constructed by SeMPLE  are called ``semi-amortized'', since an initial amortized approximation of the likelihood $p(\by|\btheta)$ for a generic $\by$ is first obtained, and then rapidly adapted to the individual-specific data $\by^{(i)}_o$, providing an approximation to the individual $p(\by^{(i)}_o|\btheta)$, without having to re-start separate fittings completely from scratch for every individual $i$, but instead starting from the amortized approximation. 

We present two versions of SeMPLE, both delivering accurate inference, as demonstrated through comparisons with exact (pseudomarginal) Bayesian inference. The first version offers greater flexibility by allowing the specification of both fixed parameters and random effects, although at a higher computational cost. In contrast, the second version is designed for enhanced scalability, but requires all parameters to be treated as random effects. To illustrate our methodology, we applied it to \textcolor{black}{three case studies based on two models}: \textcolor{black}{(i)} a mixed-effects Ornstein-Uhlenbeck state-space model, and \textcolor{black}{(ii)} a mixed-effects SDE model (used to describe translation kinetics following mRNA transfection) which is tested using both simulated and real-world data. The code is available at \url{https://github.com/henhagg/semple_mem}.

\section{Related work}\label{sec:background}

Simulation-based inference (SBI) methods, reviewed e.g. in \cite{cranmer2020frontier} and \cite{pesonen2023abc}, have also been called ``likelihood-free inference'', where the latter has been used especially with reference to approximate Bayesian computation (ABC) \citep{marin,sisson2018handbook}, synthetic likelihoods \citep{wood2010statistical,price2018bayesian}, and pseudomarginal Markov chain Monte Carlo methods \citep{andrieu2009pseudo,andrieu2010particle} when simple forward simulation is possible (as when the bootstrap filter is used to unbiasedly approximate the likelihood). The mentioned approaches have also been denoted as ``statistical SBI'' in \cite{wang2024comprehensive}, to distinguish those from more recent methods exploiting neural networks (typically normalizing flows, \citealp{rezende2015variational,papamakarios2021normalizing}) to approximate conditional densities, so-called ``neural conditional density estimation'' (NCDE), which have gained considerable attention. NCDE has been used to approximate likelihoods (\citealp{papamakarios2019,chen2021neural}), posterior distributions \citep{papamakarios2016,greenberg2019,durkan2020contrastive,chen2021neural,miller2021truncated,delaunoy2022towards}, or the likelihood and the posterior simultaneously \citep{wiqvist2021sequential,radev2023jana}. Moreover, NCDE approaches have been used both to sequentially refine inference conditionally on a specific observed data set $\by_o$, but also in an amortized way, see the review in \cite{zammit2024neural}. For amortized approaches, the trained network does not depend on any specific $\by_o$ and therefore, once training has completed, it can be used to rapidly produce conditional density estimation for any $\by_o$, though this happens at a large upfront resource investment to obtain the amortized network in the first place. 
Regarding non-amortized approaches, comparisons between some of the methods are available, e.g., in \cite{greenberg2019} and \cite{haggstrom2024}.

For the specific case of non-SBI inference for mixed-effects stochastic dynamic models, the range of inference options is large \citep{sdemem-webpage}. However, this range shrinks considerably when SBI methods are considered: for the latter, methods for mixed-effects stochastic dynamic models revolve almost exclusively around pseudomarginal Markov chain Monte Carlo (pMCMC) \citep{whitaker2017bayesian,wiqvist2021,botha2021particle,persson2022}, and an exception within SBI is the NCDE-based posterior inference in \cite{arruda2024}. The advantage of pMCMC methods is that they produce exact Bayesian inference in the limit of an infinite number of MCMC iterations. Therefore, when it is possible to use pMCMC, this provides gold-standard Bayesian inference. However, in practice, for pMCMC to be effective, advanced proposal mechanisms for the solution paths of the models are often needed, to reduce the variance of Monte-Carlo based likelihood approximations (typically via particle filters) and hence reduce the runtime to properly explore the posterior surface. In pMCMC, constructing proposals for the solution's paths is a challenging and highly-specialized task (examples are \citealp{golightly2011bayesian,del2015sequential,schauer2017guided}), and bespoke constructions often need to be produced for any different attempted model, and often depend on specific assumptions on the measurement error (e.g, a linear observation model with Gaussian measurement error). Moreover, the tuning of the parameter proposal in pMCMC (and especially its initialization) can be tedious and prone to trial-and-error. 
This is why alternative SBI methods that rely solely on ``simple'' \textit{forward} model simulation are particularly appealing, as they facilitate learning the mapping between simulated 
$\btheta$ and simulated $\by$. In our work, the goal is to construct surrogate deterministic approximations of the likelihood and posterior, rather than stochastic likelihood approximations as generated in pMCMC. In doing so, we do not employ NCDE, in contrast to \cite{arruda2024}, and instead provide a more parsimonious framework which is nevertheless expressive enough to produce accurate Bayesian inference.
Before moving further, we wish to remind the reader that in SBI it is typical to conduct inference based on summaries $S(\by)$ (provided by a data-reduction mapping) of $\by$, rather than inference based on $\by$ itself, where $S(\by)$ is a set of statistics of $\by$ that are deemed informative about $\btheta$ \citep{fearnhead2012constructing,wiqvist19a,aakesson2021convolutional} but are low-dimensional compared to $\by$. In our examples we do not employ summaries, and therefore we do not discuss this aspect further, however our methodology could accommodate inference based on some $S(\by)$ should it be necessary, and in such case all the instances where $\by$ appears could be substituted with $S(\by)$.

\section{Stochastic differential equation mixed-effects models}\label{sec:sdemem}

As mentioned in the introduction, we may consider any generative model $\mathcal{M}$ to describe time-dynamics in the data. We choose to provide illustrations based on models employing stochastic differential equations, however these could be substituted with other models, for example solvers for Markov jump processes for stochastic chemical reactions.
Consider data from a population of $M$ individuals. Assume that the dynamics for each individual $i$ are described by a stochastic process $\{\boldsymbol{X}_t^{(i)}\}_{t \geq 0}$ indexed by $t$ (where $t$ often indicates time though it can also be something different), where $\boldsymbol X^{(i)}_t \in \mathbb{R}^d$ for every $t$ and every $i=1,...,M$. Assume dynamics governed by the following stochastic differential equations (SDEs)
\begin{equation} \label{eq:sde_mem}
\begin{cases}
        d\boldsymbol{X}_t^{(i)} &= \bmu(\boldsymbol{X}_t, \bci, \bkappa, t) dt + \boldsymbol\sigma(\boldsymbol{X}_t, \bci, \bkappa, t) d\boldsymbol{B}_t^{(i)} \\ \boldsymbol{X}_0^{(i)} &= \boldsymbol{x}_0^{(i)} \quad i=1,...,M,\\
        \bc^{(i)} &\sim \pi(\bc \g \bseta),\quad i=1,...,M,
\end{cases}
\end{equation}
where $\bmu$ is a $d$-dimensional drift vector, the diffusion coefficient $\boldsymbol\sigma$ is a $d \times d$ positive definite matrix, each $\boldsymbol{B}_t^{(i)}$ denotes a vector of $d$ independent Wiener processes. In \eqref{eq:sde_mem} we assume individual-specific parameters $\bc^{(i)}\in \mathbb{R}^q$, while $\bkappa\in \mathbb{R}^p$ is common to all individuals and \textcolor{black}{$\bseta \in \mathbb{R}^u$}. For the individual-specific parameters we assume $\bc^{(i)} \sim \pi(\bc \g \bseta) \, (i = 1,\ldots, M)$ and we denote their collection across all subjects as $\bc = (\bc^{(1)}, \ldots, \bc^{(M)})$. The parameter $\bseta$ is called ``population parameter'' as it is underlying the distribution of all the $\bc^{(i)}$, and as such it does not vary with $i$. Similarly, parameter $\bkappa$ is also assumed not to  vary with $i$ and as such is common to all subjects, however, unlike $\bseta$ for the $\bc^{(i)}$'s, $\bkappa$ does not identify the distribution of any other random parameter.

The process $\{\boldsymbol{X}_t^{(i)}\}$ may be observed directly (ie without error) or indirectly: here we consider the general observational model \eqref{eq:obs_model} where it is also possible to have noisy observations $\by^{(i)}_t$ that are conditionally independent (given the latent process), and that are linked to $\{\boldsymbol{X}_t^{(i)}\}$ via
\begin{equation} \label{eq:obs_model}
    \boldsymbol{Y}_t^{(i)} = g(\boldsymbol{X}_t^{(i)}, \boldsymbol{\varepsilon}_t^{(i)}), \quad \boldsymbol{\varepsilon}_t^{(i)} \sim \pi_\varepsilon(\bxi) \quad i=1,\ldots,M,
\end{equation}
\textcolor{black}{where $\boldsymbol{Y}_t^{(i)}\in\mathbb{R}^{d_o}$, with $d_o\leq d$, $\boldsymbol{\varepsilon}_t^{(i)}\in\mathbb{R}^{d_o}$} represents  measurement errors with distribution $\pi_\varepsilon(\bxi)$ parameterised by the vector $\bxi \in \mathbb{R}^s$, and $g(\cdot)$ is a (possibly non-linear) function. We exemplify the dependence relationship between the introduced parameters and the stochastic processes in Figure \ref{fig:sdemem-network}. Evidently, if we assume no measurement error, then the observations $\by^{(i)}$ are direct (error-free) observations of $\{\boldsymbol{X}_t^{(i)}\}_{t\geq 0}$. Assume that noisy observations are collected at time-points $\{t_1,t_2,...,t_n\}$, then we can have the case where at observational time $t_j$ the observation $\by_{t_j}^{(i)}$ is a vector of length $d_o$, where having $d_o=d$ corresponds to the system being fully observed at $t_j$, whereas having $d_o<d$ opens up for $\{\boldsymbol{X}_t^{(i)}\}$ being ``partially observed''. A typical example would be $\boldsymbol{Y}_{t_j}^{(i)} = \boldsymbol{F}_{t_j}^{(i)}\boldsymbol{X}_{t_j}^{(i)}+\boldsymbol{\varepsilon}_{t_j}^{(i)}$, where the $\boldsymbol{F}_{t_j}^{(i)}$ are $d_o\times d$ matrices of known coefficients. In the notation introduced, we assumed for simplicity that the observational times are the same for all individuals, but we could have also used $\{t_1^{(i)},t_2^{(i)},...,t_{n_i}^{(i)}\}$ and assumed that the set of observations are of different lengths $n_i$ for different individuals: this can be handled in our framework but we decided to keep the notation lighter, for ease of reading. The vector of observed data for subject $i$ is therefore $\by^{(i)}_o=(\by^{(i)}_{1,o},...,\by^{(i)}_{n,o})\in\mathbb{R}^{d_o\times n}$ where we used the shorthand $\by^{(i)}_j\equiv \by^{(i)}_{t_j}$. The full set of observations stacks all individual observations as $\by_o=(\by^{(1)}_o,...,\by^{(M)}_o)^{\mathrm{T}}$. To simplify the reading, in next sections we will use $\by$ to denote a generic dataset, observed or simulated, and we will distinguish the two cases only when necessary.

Equations \eqref{eq:sde_mem}-\eqref{eq:obs_model} define a very flexible model, a \textit{stochastic differential equation mixed-effects model} (SDEMEM), which is able to represent (i) stochastic intra-individual variation (via the diffusion terms in the SDEs), (ii) between individuals variation (via the distribution  $\pi(\bc|\bseta)$ of the individual parameters) and (iii) (optional) measurement variability (via $\pi_\varepsilon(\bxi)$), if measurement error is at all considered. The model \eqref{eq:sde_mem}-\eqref{eq:obs_model} is a state-space model, namely the latent process is Markovian and observations are assumed conditionally independent given the latent process. \textcolor{black}{The Markovianity of $\{\boldsymbol{X}_t^{(i)}\}_{t\geq 0}$ implies that, in principle, exact simulation of solution paths to SDEs could arise either by sampling directly from the transition density $\boldsymbol{x}_t^{(i)}\sim \pi(\boldsymbol{x}_t^{(i)}|\boldsymbol{x}_{s}^{(i)},\bc^{(i)}, \bkappa)$, for $s<t$, or using rejection sampling approaches \citep{beskos2005exact,beskos2006exact}.}
However, typically, and except for the simplest cases, exact approaches are not feasible, and  numerical approximations  (schemes) are implemented instead. Here we consider the simplest and most commonly used approximation scheme, which is the Euler-Maruyama scheme, where the solution $\boldsymbol{x}_{t}^{(i)}$ at time $t$ is advanced to time $t+\Delta_t$ via 
\begin{align} \label{eq:em_sde}
    \boldsymbol{x}_{t + \Delta_t}^{(i)} &= \boldsymbol{x}_t^{(i)} + \bmu(\boldsymbol{x}_t^{(i)}, \bci, \bkappa, t) \Delta_t \nonumber\\
    &+ \boldsymbol\sigma(\boldsymbol{x}_t^{(i)}, \bci, \bkappa, t) \cdot \boldsymbol u_t^{(i)},
\end{align}
where $\boldsymbol u_t^{(i)} \sim \mathcal{N}(0, \Delta_t \boldsymbol I^{d \times d})$,
for some step size $\Delta_t>0$ and conditionally on $\bci$ and $\bkappa$. Here and in the following we use $\mathcal{N}(\boldsymbol{m},\boldsymbol{\Sigma})$ to denote a multivariate Gaussian distributions with mean $\boldsymbol{m}$ and covariance matrix $\boldsymbol{\Sigma}$. However, whenever possible, more accurate numerical scheme should be considered for the solution of the SDE in \eqref{eq:sde_mem} (a recent monograph is \citealp{higham2021introduction}), for example \cite{buckwar2020spectral} displays the inference bias in the posterior of the parameters when certain SDEs are solved using Euler-Maruyama. In conclusion, the generative model that in Section \ref{sec:intro} was denoted $\mathcal{M}(\btheta)$, \textcolor{black}{here it is represented by \eqref{eq:sde_mem}-\eqref{eq:obs_model}. While we could also denote the generative model with $\mathcal{M}_\Delta(\btheta)$, since a numerical scheme with stepsize $\Delta$ is required to solve the SDE, for simplicity we will keep using $\mathcal{M}(\btheta)$}. Importantly, while in this work the applications and case-studies focus on inference for SDE mixed-effects models, where observations arise from a state-space model formulation \eqref{eq:sde_mem}-\eqref{eq:obs_model}, this is only a possible example of application of the methodology offered in Sections \ref{sec:semple_gllim}-\ref{sec:semple_mem}.
\textcolor{black}{In fact, in \cite{haggstrom2024} the original SeMPLE methodology, which was not specialized to mixed-effects state-space SDE models,  was tested on examples that included both static and dynamic models, the only requirement being the ability to simulate data from a generative model $\mathcal{M}(\btheta)$. Therefore, in this work we confirm that the methodology in  \cite{haggstrom2024} is flexible. As an additional possibility (not pursued in this work), it would not be difficult to show the possibility to accommodate, for example, observations arising from  non-Markovian processes $\{X_t^{(i)}\}_{t\geq 0}$, or where the $Y_t^{(i)}|X_t^{(i)}$ are not conditionally independent.}

\begin{figure*}
    \centering
    \begin{tikzpicture}[
    box/.style={draw, minimum size=1cm, rounded corners, text centered, font=\small},
    arrow/.style={-Latex, thick},
    node distance=1.5cm and 3.5cm
    ]
    
    \node[box, label=above:Population parameters] (eta) {$\eta$};
    \node[box, right=of eta, label=above:Common parameters] (kappa) {$\kappa$};
    \node[box, right=of kappa, label={[align=center]above:Measurement noise \\ parameters}] (xi) {$\xi$};
    \node[box, below=of eta, label=below:Individual parameters] (ci) {$c^{(i)}$};
    \node[box, right=of ci, label=below:Latent process] (x) {$X_t^{(i)}$};
    \node[box, right=of x, label=below:Observable process] (yi) {$Y_t^{(i)}$};
    
    \draw[dotted, thick] ($(ci.north west)+(-2,0.5)$) rectangle ($(yi.south east)+(1.5,-1)$);
    \node at ($(ci.north east)+(2,0.2)$) {$i = 1, \dots, M$};
    
    \draw[arrow] (eta) -- (ci);
    \draw[arrow] (ci) -- (x);
    \draw[arrow] (x) -- (yi);
    \draw[arrow] (kappa) -- (x);
    \draw[arrow] (xi) -- (yi);
    
    \end{tikzpicture}
    \caption{Bayesian network model structure for a SDE mixed-effects model.}
    \label{fig:sdemem-network}
\end{figure*}

\section{Bayesian inference for SDEMEMs} \label{sec:bayesian_inference}

As mentioned in the previous section, we will use $\by$ to denote both simulated data and observed data, and we will make use of the notation $\by_o$ for observed data only when necessary.
Denoting with $\by = (\by^{(1)}, \ldots, \by^{(M)})^{\mathrm{T}}$ the stacked data from all $M$ individuals, the full vector of parameters to learn is $\btheta=(\bc, \bkappa, \bseta, \bxi)$, where $\bc=(\bc^{(1)},...,\bc^{(M)})$. Notice, if data are not assumed to be affected with measurement error, then $\bxi$ is not included in $\btheta$. Below we consider the most general scenario. Assume that the $\bc^{(i)}$'s are mutually independent and that data $\by^{(i)}$ are independent conditionally on the $\bc^{(i)}$'s.
We wish to sample from the full posterior distribution
\begin{align} \label{eq:full_post}
    &\pi(\bc, \bkappa, \bseta, \bxi \g \by) \propto \nonumber \\ 
    &\pi(\bkappa, \bseta, \bxi) \prod_{i=1}^{M} \pi(\by^{(i)} \g \bc^{(i)}, \bkappa, \bxi) \pi(\bc^{(i)} | \bseta),
\end{align}
where the individual likelihoods are obtained by the marginalization
\begin{align}
    \pi(\by^{(i)} &\g \bc^{(i)}, \bkappa, \bxi) = \int \pi(\by^{(i)}, \boldsymbol x^{(i)} \g \bc^{(i)}, \bkappa, \bxi) d\boldsymbol x^{(i)} \label{eq:likelihood_marginalize_x} \\ 
   & = \int \prod_{j=1}^{n} \pi(\by_j^{(i)} | \boldsymbol x_j^{(i)}, \bxi) \pi(\boldsymbol x_0^{(i)}|\bc^{(i)}, \bkappa) \label{eq:likelihood_integral} \times \nonumber \\
    &\prod_{j=1}^{n} \pi(\boldsymbol x_j^{(i)} \g \boldsymbol x_{j-1}^{(i)}, \bc^{(i)}, \bkappa) d\boldsymbol x_0^{(i)} \cdots d\boldsymbol x_{n}^{(i)},
\end{align}
where the $\boldsymbol x_j^{(i)}\equiv \boldsymbol x_{t_j}^{(i)}$ are values of the $\{\boldsymbol X^{(i)}_t\}_{t\geq 0}$ process at the observational times, and
where $\boldsymbol x_{0}^{(i)}$ is a starting value for $\{\boldsymbol X^{(i)}_t\}_{t\geq 0}$ at time $t=0$. The integral \eqref{eq:likelihood_integral} is generally analytically intractable, except for simple cases, which motivates the need for either particle methods (sequential Monte Carlo) to get an unbiased estimate of the individual likelihood or, as we propose, a simulation-based method that provides a deterministic surrogate model as an approximation of the individual likelihood.

\textcolor{black}{The graphical dependency in Figure \ref{fig:sdemem-network}, coupled with the state-space structure of the model (equations \eqref{eq:sde_mem}-\eqref{eq:obs_model}), suggests a natural factorization of the full posterior \eqref{eq:full_post}, from which we sample using a Gibbs sampler, following \cite{wiqvist2021}, \cite{botha2021particle}, \cite{persson2022}, who used the same factorization for SDEMEMs. The Gibbs sampler iterates through the following steps}
\small
\begin{align}
    \text{step 1: \,}\pi(\bc \g \bkappa, \bseta, \bxi, \by) & \propto \prod_{i=1}^{M} \pi(\bc^{(i)} \g \bseta) \, \pi(\by^{(i)} \g \bc^{(i)}, \bkappa, \bxi) \label{eq:gibbs_c} \\
    \text{step 2: \,}\pi(\bkappa, \bxi \g \bc, \bseta, \by) & \propto \pi(\bkappa, \bxi) \prod_{i=1}^{M} \pi(\by^{(i)} \g \bc^{(i)}, \bkappa, \bxi) \label{eq:gibbs_kappaxi} \\
    \text{step 3: \,}\pi(\bseta \g \bc, \bkappa, \bxi, \by) & \propto \pi(\bseta) \prod_{i=1}^{M} \pi(\bc^{(i)} \g \bseta). \label{eq:gibbs_eta}
\end{align}
\normalsize
Notice that the first step (equation \eqref{eq:gibbs_c}) is equivalent to \eqref{eq:gibbs_ind_c}, which in practice we employ in our implementation. That is, since we assume that data from each individual $i$ are conditionally independent from the data of another individual $i'$ ($i'\neq i$), then we can separately sample each individual $\bc^{(i)}$ ($i = 1,...,M$) from 
\begin{equation}
    \pi(\bc^{(i)} \g \bkappa, \bseta, \bxi, \by^{(i)}) \propto \pi(\bc^{(i)} \g \bseta) \, \pi(\by^{(i)} \g \bc^{(i)}, \bkappa, \bxi), \label{eq:gibbs_ind_c}
\end{equation}
then stack the obtained draws into $\bc$. The fact that the first step is broken-down into $M$ independent contributions allows for parallelization, and in case of a large number of individuals this could reduce the runtime significantly.
The second step in the Gibbs sampler pertains to the sampling of fixed effect parameters $\bkappa$ and $\bxi$, which are shared by all individuals, \textcolor{black}{and the fact that their sampling is separated from that of the $\bc^{(i)}$'s  allows $(\bkappa, \bxi)$ to be treated as common across all individuals}. Notice that our treatment is different from, for example, \cite{arruda2024} where such shared parameters \textcolor{black}{are modeled as random effects with zero variance.}
A good reason for considering a Gibbs sampler is that its first step \eqref{eq:gibbs_c} (or equivalently \eqref{eq:gibbs_ind_c}), which requires a Metropolis-within-Gibbs approach, can benefit from an automatically learned proposal sampler that is particularly suited for multimodal targets, as we detail in Section \ref{sec:semple_gllim}. 
However, in practice, sampling from  \eqref{eq:gibbs_c}-\eqref{eq:gibbs_kappaxi} when dealing with SDEMEMs is difficult, in that the individual likelihoods $\pi(\by^{(i)} \g \bc^{(i)}, \bkappa, \bxi)$  are typically intractable large-dimensional integrals (see \eqref{eq:likelihood_integral}) that get approximated via Monte Carlo simulations, and we refer to \cite{wiqvist19a}, \cite{botha2021particle} and \cite{persson2022} for powerful but computer intensive approaches based on particle filters. Even worse, these computationally intensive Monte Carlo approximations, resulting in pseudomarginal pMCMC methods, are difficult to tune and have to be re-executed for any considered $\btheta$ and whenever a new $\by^{(i)}$ is considered. Hence large values of individuals ($M$) can break down the feasibility of such approaches (but see the ameliorating strategies in \citealp{persson2022}). 
To address this problem, in the next section, we show how to obtain both an amortized surrogate likelihood model, and an amortized surrogate posterior model in the form of two Gaussian mixture models (``mixture of experts''). Moreover, we anticipate that there can be computational advantages in eliminating the second Gibbs step (hence assuming that all parameters are random effects), which is discussed in Section \ref{sec:scalable-semple}.

\section{Likelihood and posterior estimation via mixtures of experts} \label{sec:semple_gllim}
The main contribution of this work is to create a framework for Bayesian inference where the likelihood $\pi(\by \g \bc, \bkappa, \bxi)$ is first approximated in an amortized way and then is specialized to a given observation $\by_o^{(i)}$ (for every $i=1,...,M$). 
We say that our method is ``semi-amortized'' because, unlike amortized procedures, we do not simply plug the observation $\by_o^{(i)}$ into an amortized approximation of $\pi(\by \g \bc, \bkappa, \bxi)$ that can be evaluated, but instead start from an amortized likelihood, and this is sequentially refined for the specific observation $\by_o^{(i)}$, to provide the final $\pi(\by_o^{(i)} \g \bc^{(i)}, \bkappa, \bxi)$.
We will see that the same computations involved in producing the approximation to $\pi(\by_o^{(i)} \g \bc^{(i)}, \bkappa, \bxi)$ will return, as a by-product and without any further computation, an approximation to the posterior $\pi(\bc^{(i)}, \bkappa, \bxi\g \by_o^{(i)})$.
This is achieved via the \textit{SEquential Mixtures Posterior and Likelihood Estimation} (SeMPLE) inference of \cite{haggstrom2024}. Once all individual likelihoods $\pi(\by_o^{(i)} \g \bc^{(i)}, \bkappa, \bxi)$ are approximated with \textit{deterministic} surrogate functions, they are used in place of intractable likelihoods in \eqref{eq:gibbs_c} and \eqref{eq:gibbs_kappaxi}. This is different from the pseudomarginal MCMC approach in \cite{wiqvist2021}, \cite{botha2021particle} and \cite{persson2022}, that compute Monte Carlo estimates of the likelihoods $\pi(\by^{(i)} \g \bc^{(i)}, \bkappa, \bxi)$. It is of interest to investigate whether the surrogate likelihoods from SeMPLE provide accurate approximations of the true intractable likelihood and, thus, accurate posterior inference. It is also of interest to discuss the computational footprint of using SeMPLE to estimate the likelihoods, compared to gold-standard pseudomarginal methods for SDEMEMs.

The remaining of this section (up until equation \eqref{eq:gibbs_gllim_c}) is not specific to stochastic mixed effects models, and here we use $\by$ to denote a generic observable \textcolor{black}{and $\btheta$ to denote a generic model parameter}.
SeMPLE uses surrogates that are mixture-of-experts whose parameters are fitted via expectation-maximization \citep{xu1994alternative} on training data obtained via an efficient Markov chain Monte Carlo (MCMC) algorithm. The SeMPLE approach makes repeated use, in a sequential way, of Gaussian locally linear mappings (GLLiM, \citealp{deleforge2014}). GLLiM introduces a mixture of experts in the form of a GMM defined on the joint distribution for $(\btheta, \by)$. The relationship between $\by \in \mathbb{R}^{d_o \times n}$ and $\btheta \in \mathbb{R}^l$ is assumed to be locally linear, defined using a latent variable $z \in \{1,...,K\}$, via the following \textit{surrogate generative model}
\begin{equation}
    \by = \sum_{k=1}^K \mathbb{I}_{\{z = k\}} (\tilde{\boldsymbol A}_k \btheta + \tilde{\boldsymbol b}_k + \tilde{\boldsymbol \epsilon}_k),\label{eq:affine}
\end{equation}
where  $\mathbb{I}$ denotes the indicator function, and $\tilde{\boldsymbol A}_k\in\mathbb{R}^{(d_o \times n)\times l}$  and $\tilde{\boldsymbol b}_k\in\mathbb{R}^{d_o \times n}$ are matrices and vectors, respectively, defining the affine transformation of $\btheta$ in \eqref{eq:affine}, while $\tilde{\boldsymbol \epsilon}_k\in\mathbb{R}^{d_o \times n}$ corresponds to an error term capturing both the observational
noise and the modelling error due to assuming an affine approximation for the data. In the following we consider  Gaussian noise, 
$\tilde{\boldsymbol \epsilon}_k \sim \mathcal{N}(\boldsymbol 0, \tilde{\boldsymbol \Sigma}_k$), and assume $\tilde{\boldsymbol \epsilon}_k$ to not depend on $\btheta$, $\by$ nor $z$.

Conditionally on component $z=k$, an approximate generative model is given by
\begin{equation}
    q_{\tilde{\bphi}}(\by \g \btheta, z=k) = \mathcal{N}(\by; \tilde{\boldsymbol A}_k \btheta + \tilde{\boldsymbol b}_k, \tilde{\boldsymbol \Sigma}_k).
\end{equation}
To complete the hierarchical model, $\btheta$ is assumed to follow a mixture of Gaussian distributions specified by
\begin{equation}
     q_{\tilde{\bphi}}(\btheta \g z=k) = \mathcal{N}_l(\btheta; \tilde{\boldsymbol \nu}_k, \tilde{\boldsymbol \Gamma}_k), \quad
     q_{\tilde{\bphi}}(z=k) = \pi_k. \label{eq:pik}
\end{equation}
The GLLiM hierarchical construction above  (eq.(\ref{eq:affine}) to (\ref{eq:pik})) defines a joint  GMM on $(\by,\btheta)$:
\begin{equation*}
     q_{\tilde{\bphi}}(\by, \btheta)\! = \!\sum_{k=1}^K  q_{\tilde{\bphi}}(\by \!\g\! \btheta, z\!=\!k)  q_{\tilde{\bphi}}(\btheta \!\g\! z\!=\!k)  q_{\tilde{\bphi}}(z\!=\!k),
\end{equation*}
where the full vector of mixture model parameters is
$
    \tbphi = \{\pi_k, \tilde{\boldsymbol{\nu}}_k, \tilde{\boldsymbol \Gamma}_k, \tilde{\boldsymbol A}_k, \tilde{\boldsymbol b}_k, \tilde{\boldsymbol \Sigma}_k\}_{k=1}^K.
$
Given this joint distribution, we can easily deduce the conditional distributions $q_{\tilde \bphi}(\by \g \btheta)$ and $q_{\bphi}(\btheta \g \by)$ in closed-form. First, we have the surrogate likelihood
\begin{equation}
    \label{eq:surr_lik}
    q_{\tilde{\bphi}}(\by \g \btheta) = \sum_{k=1}^{K} \tilde{\omega}_k(\btheta) \mathcal{N}(\by; \tilde{\boldsymbol{A_k}} \btheta + \tilde{\boldsymbol{b}}_k, \tilde{\boldsymbol{\Sigma}}_k),
\end{equation}
with
\begin{equation}
    \label{eq:lik_prob}
    \tilde{\omega}_k(\btheta) = \frac{\pi_k \mathcal{N}_l(\btheta; \tilde{\boldsymbol{\nu}}_k, \tilde{\boldsymbol{\Gamma}}_k)}{\sum_{j=1}^{K} \pi_j \mathcal{N}_l(\btheta; \tilde{\boldsymbol{\nu}}_j, \tilde{\boldsymbol{\Gamma}}_j)}.
\end{equation}
Notice the important distinction that $q_{\tilde{\bphi}}(\by \g \btheta)$ is the approximate (surrogate) likelihood, while $\pi(\by|\btheta)$ is the true (unknown) likelihood that underlines the generative model that we denoted with $\mathcal{M}(\btheta)$ in Section \ref{sec:intro} and that, for SDEMEMs, corresponds to equations \eqref{eq:sde_mem}-\eqref{eq:obs_model}.
\textcolor{black}{As a consequence of assuming a mixture model as the joint model for $(\by,\btheta)$, we immediately obtain also a surrogate posterior, without the need for further training}, and this posterior is given by
\begin{equation}
    \label{eq:surr_post}
    q_{{\bphi}}(\btheta \g \by) = \sum_{k=1}^{K} \omega_k(\by) \mathcal{N}_l(\btheta; \boldsymbol A_k \by + \boldsymbol b_k, \boldsymbol \Sigma_k),
\end{equation}
with
\begin{equation}
    \label{eq:post_prob}
    \omega_k(\by) = \frac{\pi_k \mathcal{N}(\by; \boldsymbol \nu_k, \boldsymbol \Gamma_k)}{\sum_{j=1}^{K} \pi_j \mathcal{N}(\by; \boldsymbol \nu_j, \boldsymbol \Gamma_j)},
\end{equation}
where
$\bphi = \{\pi_k, \boldsymbol \nu_k, \boldsymbol \Gamma_k, \boldsymbol A_k, \boldsymbol b_k, \boldsymbol \Sigma_k\}_{k=1}^K$. The surrogate \eqref{eq:surr_post} is obtained instantaneously and in closed-form, because we have the following algebraic relationships deduced from $\tilde{\bphi}$   \citep{deleforge2014}: 
\begin{align} 
    \boldsymbol{\nu}_k &= \tilde{\boldsymbol A_k} \tilde{\boldsymbol \nu}_k + \tilde{\boldsymbol b}_k, \label{eq:forwardParam_line0} \\
    \boldsymbol \Gamma_k &= \tilde{\boldsymbol \Sigma}_k + \tilde{\boldsymbol A_k} \tilde{\boldsymbol \Gamma}_k \tilde{\boldsymbol A_k}^\top, \label{eq:forwardParam_line1} \\
    \boldsymbol \Sigma_k &= \big(\tilde{\boldsymbol \Gamma}_k^{-1} + \tilde{\boldsymbol A_k}^\top \tilde{\boldsymbol \Sigma}_k^{-1} \tilde{\boldsymbol A_k} \big)^{-1}, \\
    \boldsymbol A_k &= \boldsymbol \Sigma_k \tilde{\boldsymbol A_k}^\top \tilde{\boldsymbol \Sigma}_k^{-1}, \label{eq:forwardParam_line2}\\
    \boldsymbol b_k &= \boldsymbol \Sigma_k \big(\tilde{\boldsymbol \Gamma}_k^{-1} \tilde{\boldsymbol{\nu}}_k - \tilde{\boldsymbol A_k}^\top \tilde{\boldsymbol \Sigma}_k^{-1} \tilde{\boldsymbol b}_k \big).\label{eq:forwardParam_line3}
\end{align}
Therefore, given training data of model parameters and simulated data $\{\btheta_j, \by_j\}_{j=1}^N$, $\tilde{\bphi}$ is first estimated via expectation-maximization (EM, \citealp{xu1994alternative,deleforge2014}), to produce $q_{\tilde \bphi}(\by \g \btheta)$, and then the corresponding approximation for ${\bphi}$ is obtained at no cost via \eqref{eq:forwardParam_line0}--\eqref{eq:forwardParam_line3}, so that the surrogate posterior $q_{\bphi}(\btheta \g \by)$ is also identified. To run EM, our implementation of SeMPLE uses the R \texttt{xLLiM} package \citep{xllim}, which gives several options to parametrize the covariance matrices $\tilde{\boldsymbol{\Sigma}}_k$ (and hence, implicitly, also the ${\boldsymbol{\Sigma}}_k$), see Appendix \ref{sec:bic} for more details.  Before moving forward, notice that the number of components $K$ used either in \eqref{eq:surr_lik} or \eqref{eq:surr_post} can be set in a principled way, as described in Appendix \ref{sec:bic}, where the Bayesian information criterion (BIC) is employed. 

Rewriting the Gibbs sampler in equations \eqref{eq:gibbs_c}-\eqref{eq:gibbs_eta} with the surrogate likelihood $q_{\tilde \bphi}$ results in the following approximated Gibbs steps (the third step is still exact as it does not involve likelihoods)
\begin{align}
    \hat{\pi}(\bc^{(i)} \g \bkappa, \bseta, \bxi, \by^{(i)}) & \propto\pi(\bc^{(i)} \g \bseta) \, q_{\tilde{\phi}}(\by^{(i)} \g \bc^{(i)}, \bkappa, \bxi),\label{eq:gibbs_gllim_c}\\ 
    &i=1,...,M. 
\nonumber \\
    \hat{\pi}(\bkappa, \bxi \g \bc, \bseta, \by) & \propto \pi(\bkappa, \bxi) \prod_{i=1}^{M}{q_{\tilde\bphi}(\by^{(i)} \g \bc^{(i)}, \bkappa, \bxi)} \label{eq:gibbs_gllim_kappaxi} \\
    \pi(\bseta \g \bc, \bkappa, \bxi, \by) & \propto \pi(\bseta) \prod_{i=1}^{M} \pi(\bc^{(i)} \g \bseta). \label{eq:gibbs_gllim_eta}
\end{align}
Note that, in terms of the GLLiM notation, we have $\btheta = (\bc, \bkappa, \bxi)$ rather than $\btheta = (\bc, \bkappa, \bseta, \bxi)$, and this is because, since the third Gibbs step does not involve likelihoods, GLLiM is not used to infer $\bseta$, given that $\bseta$ can be sampled in \eqref{eq:gibbs_gllim_eta} in an exact way, either via conjugacy or using of-the-shelves algorithms such as the No-U-Turn Sampler (NUTS, \citealp{hoffman2014no}), \textcolor{black}{which we employed via the probabilistic framework \texttt{Stan} \citep{Stan}, using default settings}. As the surrogate likelihood provided by GLLiM is a mixture of Gaussians, this is differentiable, and therefore, it is possible to use gradient-informed MCMC methods such as NUTS, as implemented in \texttt{Stan}. \textcolor{black}{Compared to the previously mentioned pseudo-marginal method found in \cite{persson2022} and named \textcolor{black}{PEPSDI (standing for \texttt{P}articles \texttt{E}ngine for \texttt{P}opulation \texttt{S}tochastic \texttt{D}ynam\texttt{I}cs)}, which requires non-trivial tuning of the covariance matrix of the proposal kernel, we found that with \texttt{Stan} the implemented warm-up scheme worked well without user input, when provided with a reasonable start-guess for $\btheta$.}

However, as we motivate in Section \ref{sec:semple_mem}, we will prefer to employ $q_{\bphi}(\btheta|\by)$ as a multimodal proposal sampler in step 1 (eq. \eqref{eq:gibbs_gllim_c}), and instead employ HMC only for the other two steps. Furthermore, in Section \ref{sec:scalable-semple}, we illustrate another, more scalable, Gibbs sampler associated to assuming all parameters as random effects.
To efficiently apply this sampler with the surrogate likelihood, we first train the GLLiM model on simulated data in a sequential manner, ensuring that the data become increasingly informative for the observed $\by_o$. The full procedure is described in the SeMPLE mixed-effects algorithm in the following section.

\begin{algorithm*}[tp]
 \small
    \caption{SeMPLE for mixed-effects models}
    \label{alg:unperturbed}
    \hspace*{\algorithmicindent} \textbf{Input:} $\pi(\bc\g\bseta)$, $\pi(\bseta)$, $\pi(\bkappa, \bxi)$, observed data $\{\by_o^{(i)}\}_{i=1}^{M}$, simulator $\pi(\by \g \bc, \bkappa, \bxi)$, positive integers $R$ (number of SeMPLE rounds), $N$ (number of prior predictive simulations at round $r=0$), $N_g$ (number of Gibbs samples).\\
    \hspace*{\algorithmicindent} \textbf{Output:} Posterior samples $\{\{\bc_j^{(i)}\}_{i=1}^M, \bkappa_j, \bxi_j, \bseta_{j}\}_{j=1}^{N_g}$\\
    \begin{algorithmic}[1]
        \item[]
        \textbf{Round $r=0$}
        \State Sample iid $\bseta_j \sim \pi(\bseta)$, $j = 1,...,N$ \label{alg:sample_prior_eta}
        \State Sample  $\bc_j \sim \pi(\bc \g \bseta_j)$, $j = 1,...,N$
        \State Sample  $(\bkappa_j, \bxi_j) \sim \pi(\bkappa, \bxi)$, $j = 1,...,N$
        \State Simulate $N$ single-individual datasets $\by^{(j)} \sim \pi(\by \g \bc_j, \bkappa_j, \bxi_j)$, $j = 1,...,N$ \label{alg:sim_data_prior}
        \State Collect $\mathcal{D}_0 = \{(\bc_j, \bkappa_j, \bxi_j), \by_j\}_{j=1}^{N}$
        \State Train \textcolor{black}{single individual} likelihood $q_{\tilde{\bphi}_0}(\by \g \bc, \bkappa, \bxi)$ on $\mathcal{D}_0$. \label{alg:train_likelihood_prior}
        \State Obtain $q_{\bphi_0}(\bc, \bkappa, \bxi \g \by)$ \textcolor{black}{ in closed form from $q_{\tilde{\bphi}_0}(\by \g \bc, \bkappa, \bxi)$,  using \eqref{eq:forwardParam_line0}--\eqref{eq:forwardParam_line3}}. \label{alg:train_post_prior}\\

        \item[]
        \textbf{Round $r=1$}
        \For{$i = 1:M$}
            \State Sample from surrogate posterior iid $(\bc_j^{(i)}, \bkappa_j^{(i)}, \bxi_j^{(i)}) \sim q_{\bphi_0}(\bc, \bkappa, \bxi \g \by_o^{(i)})$, $j=1,...,\textcolor{black}{N/M}$ \label{alg:sample_surrogate_posterior}
            \State Simulate $\by_j^{(i)} \sim \pi(\by \g \bc_j^{(i)}, \bkappa_j^{(i)}, \bxi_j^{(i)})$, $j=1,...,N/M$ \label{alg:simulate_surrogate_posterior}
            \State Collect  $\mathcal{D}_1^{i} = \{(\bc_j^{(i)}, \bkappa_j^{(i)}, \bxi_j^{(i)}), \by_j^{(i)}\}_{j=1}^{N/M}$
        \EndFor
        \State Train \textcolor{black}{single individual} likelihood $q_{\tilde\bphi_1}(\by \g \bc, \bkappa, \bxi)$ on $\mathcal{D}_1 = \bigcup_{i=1}^{M} \mathcal{D}_1^{i}$ \label{alg:train_likelihood_surrpost}
        \State Obtain $q_{\bphi_1}(\bc, \bkappa, \bxi \g \by)$ \textcolor{black}{ in closed form from $q_{\tilde{\bphi}_0}(\by \g \bc, \bkappa, \bxi)$,  using \eqref{eq:forwardParam_line0}--\eqref{eq:forwardParam_line3}}.  \label{alg:train_post_surrpost}
        
        \item[]
        \State Initialize $\bc_1 \leftarrow \bc_{N/M}$, $(\bkappa_1, \bxi_1) = (\bar\bkappa_{N/M}, \bar\bxi_{N/M})$, and $\bseta_1 \sim \pi(\bseta)$ \label{alg:init_gibbs}
        \For{$r = 2:R$}
            \For{$j = 2:N_g$}
                \State Sample new $\bc_j^{(i)} \sim \pi(\bc_{j-1}^{(i)} \g \bseta_{j-1}) q_{\tilde\bphi_{r-1}}(\by_o^{(i)} \g \bc_{j-1}^{(i)}, \bkappa_{j-1}, \bxi_{j-1})$, $i=1,...,M$ via Algorithm \ref{alg:MH_c_unperturbed} \label{alg:gibbs_c}
                \State Sample new $(\bkappa_j, \bxi_j) \sim \pi(\bkappa_{j-1}, \bxi_{j-1}) \prod_{i=1}^{M} q_{\tilde\bphi_{r-1}}(\by_o^{(i)} \g \bc_{j}^{(i)}, \bkappa_{j-1}, \bxi_{j-1})$ via HMC \label{alg:gibbs_kappa_xi}
                \State Sample new $\bseta_j \sim \pi(\bseta_{j-1}) \prod_{i=1}^{M} \pi(\bc_{j}^{(i)} \g \bseta_{j-1})$ via HMC or conjugacy\label{alg:gibbs_eta}
                \State Simulate $\by_j^{(i)} \sim \pi(\by \g \bc_j^{(i)}, \bkappa_j, \bxi_j)$, $i=1,...,M$ \label{alg:gibbs_simulate}
            \EndFor
        \State Collect $\mathcal{D}_r = \Big\{\{\bc_j^{(i)}\}_{i=1}^M, \bkappa_j, \bxi_j, \{\by_j^{(i)}\}_{i=1}^M\Big\}_{j=1}^{N_g}$
        \State Train \textcolor{black}{single individual} likelihood $q_{\tbphi_r}(\by \g \bc, \bkappa, \bxi)$ on $\bigcup_{\tilde r=1}^{r} \mathcal{D}_{\tilde r}$ \label{alg:train_likelihood_gibbs}
        \State Obtain $q_{\bphi_r}(\bc, \bkappa, \bxi \g \by)$ \textcolor{black}{ in closed form from $q_{\tilde{\bphi}_0}(\by \g \bc, \bkappa, \bxi)$,  using \eqref{eq:forwardParam_line0}--\eqref{eq:forwardParam_line3}}. \label{alg:train_post_gibbs}
        \EndFor
        \end{algorithmic}
\end{algorithm*}

\begin{algorithm*}[htbp]
\small
    \caption{\textcolor{black}{Independence-Metropolis-Hastings for $\bc$}}
    \label{alg:MH_c_unperturbed}
    \begin{algorithmic}[1]
        \item[] Here round $r$ is $r\geq 2$ and  $q_{\bphi_{r-1}}(\bc \g \by_o^{(i)})$ produces independent samples. For the initial value $\tilde\bc_1^{(i)}$, pick the last value $\bc_{j-1}^{(i)}$ stored in the Markov chain from the previous Gibbs run.
            \State Propose $\tilde\bc_*^{(i)} \sim q_{\bphi_{r-1}}(\bc \g \by_o^{(i)})$ independently,
            \State $\alpha = \min\biggl\{1, 
            \frac{\pi(\tilde\bc_*^{(i)} \g \bseta_{j-1}) q_{\tilde\bphi_{r-1}}(\by_o^{(i)}\g \tilde\bc_*^{(i)}, \bkappa_{j-1}, \bxi_{j-1})} 
            {\pi(\tilde\bc_{1}^{(i)} \g \bseta_{j-1}) q_{\tilde\bphi_{r-1}}(\by_o^{(i)} \g \tilde\bc_{1}^{(i)}, \bkappa_{j-1}, \bxi_{j-1})} \times
            \frac{q_{\bphi_{r-1}}(\tilde\bc_{1}^{(i)} \g \by_o^{(i)})}{q_{\bphi_{r-1}}(\tilde\bc_*^{(i)} \g \by_o^{(i)})}\biggr\}$
            \State Sample $u \sim \mathcal{U}[0,1]$
            \If{$u \leq \alpha$}
                \State $\tilde\bc_2^{(i)} = \tilde\bc_*^{(i)}$ 
            \Else
                \State $\tilde\bc_2^{(i)} = \tilde\bc_{1}^{(i)}$
            \EndIf
        \State $\bc_{j}^{(i)} = \tilde\bc_2^{(i)}$
    \end{algorithmic}
\end{algorithm*}

\section{SeMPLE for mixed-effects models} \label{sec:semple_mem}

\textcolor{black}{Now that we have outlined (in Section \ref{sec:semple_gllim}) the flexible structure of SeMPLE for a generic generative model  $\mathcal{M}$, giving rise to simulated observations $\by$ or actual observations $\by_o$, we now specialize it to the case where $\mathcal{M}$ is a state-space model with mixed-effects. Therefore, by recalling Section \ref{sec:sdemem}, we now have to consider that $\by_o$ consists of a collection of observations $\by^{(i)}$ across $M$ individuals, hence $\by_o=(\by^{(1)},...,\by^{(M)})$, that individual data $\by^{(i)}$ are conditionally independent given random effects $\bc^{(i)}$, and that for each individual the observed data $\by^{(i)}_{t}$, at time $t$, are considered as noisy observations of a latent process $\boldsymbol{X}_{t}^{(i)}$ at time $t$.} 

\textcolor{black}{Our inference approach alternates between the following two tasks: (i) \textbf{sampling} via Gibbs a batch of size $N_g$ (or $N$ depending on cases) of parameters $\btheta=(\bc,\bkappa,\bseta,\bxi)$ from (an approximation of) the full posterior, and (ii) \textbf{training} the surrogate posterior and surrogate likelihood by fitting a dataset $\mathcal{D}=\{\btheta_j,\by_j\}_{j=1}^{N_g\text{ or }N}$ using expectation-maximization (via GLLiM, as introduced in Section \ref{sec:semple_gllim}), where $\btheta_j\rightarrow \mathcal{M}(\btheta_j)\rightarrow \by_j$. The training step (ii) is repeated for $R$ ``rounds'', each producing increasingly more accurate approximations of the likelihood and the posterior for the given observed data, and the latter approximations are then used in the sampling step, to provide posterior samples of increasing quality. The cycle alternating between sampling and training is repeated $R$ times, and the samples obtained at the $R$-th round provide the final (SeMPLE) inference. In the following, for simplicity of notation, we assume $N$ to be an integer multiple of $M$, so that $N/M$ is integer.  
}

The SeMPLE inference scheme for mixed-effects models is described in Algorithm \ref{alg:unperturbed}. We will see that the surrogate likelihood and the surrogate posterior are initially amortized, and then become specialized to the specific observed data $\by_o$.
At first, in round $r=0$, an initial batch of $N$ tuples $\mathcal{D}_0 = \{(\bc_j, \bkappa_j, \bxi_j), \by_j\}_{j=1}^{N}$ are produced from the prior-predictive distributions (steps \ref{alg:sample_prior_eta}-\ref{alg:sim_data_prior} of Algorithm \ref{alg:unperturbed}). \textcolor{black}{Notice, here each simulated $\by_j\in\mathbb{R}^{d_o\times n}$ has the same dimension of data from a single individual $\by^{(i)}_o$, and not the dimension $\mathbb{R}^{(d_o\times n)\times M}$ of the full dataset $\by_o=\{\by_o^{(i)}\}_{i=1}^M$, that is, $\by_j$ is not a collection of $M$ simulated individual datasets. Similarly, each $\bc_j\in\mathbb{R}^u$ inside $\mathcal{D}_0$ has the same dimension of an ``individual $\bc^{(i)}$''.} The set $\mathcal{D}_0$ constitutes the initial training data for GLLiM, and \textcolor{black}{with this training data we apply the methodology in Section \ref{sec:semple_gllim} to obtain an \textit{amortized} surrogate likelihood  $q_{\tbphi_0}(\by \g \bc, \bkappa, \bxi)$ (step \ref{alg:train_likelihood_prior}) with input argument $\by\in\mathbb{R}^{d_o\times n}$, that is a generic \textit{single-individual} data. This is a key aspect of our methodology, because after training we can plug observations from any individual $\by\leftarrow \by_o^{(i)}$ into $q_{\tbphi_0}(\by \g \bc, \bkappa, \bxi)$, and therefore immediately evaluate each individual likelihood $q_{\tbphi_0}(\by^{(i)}_o \g \bc, \bkappa, \bxi)$ at observed data $\by_o^{(i)}$. For extra clarity, it is useful to remark once more that in $\mathcal{D}_0$ we have stacked many individual $\{\bc_j\}_{j=1}^N$ parameters, with $\dim(\bc_j)=\dim(\bc^{(i)})=u$, and therefore, with some abuse of notation\footnote{In fact, here $\bc$ is to be interpreted as a $u$-dimensional real variable, and not as the collection of $M$ individual parameters $ (\bc^{(1)},...,\bc^{(M)})$ that we defined in section \ref{sec:sdemem}.}, the dimension of the $\bc$ that we write inside $q_{\tbphi_0}(\by \g \bc, \bkappa, \bxi)$ and in $q_{\bphi_0}(\bc, \bkappa, \bxi \g \by)$ (step \ref{alg:train_post_prior}), is the same dimension of an ``individual $\bc^{(i)}$'', and this feature is preserved in the next rounds of SeMPLE.} In steps \ref{alg:train_likelihood_prior}-\ref{alg:train_post_prior} of Algorithm \ref{alg:unperturbed} the mixture model parameters $\tbphi_0$  are obtained (via EM within GLLiM, as in Section \ref{sec:semple_gllim}) to produce the initial amortized \textcolor{black}{ single-individual likelihood} $q_{\tbphi_0}(\by \g \bc, \bkappa, \bxi)$, \textcolor{black}{and then by using using \eqref{eq:forwardParam_line0}--\eqref{eq:forwardParam_line3} we immediately deduce $\bphi_0$, and hence the corresponding amortized  surrogate posterior $q_{\bphi_0}(\bc, \bkappa, \bxi \g \by)$ where, again, $\by\in\mathbb{R}^{d_o\times n}$ has the dimensions of single-individual data.} This concludes round $r=0$ of SeMPLE.

Next comes round $r=1$, where steps \ref{alg:sample_surrogate_posterior}-\ref{alg:simulate_surrogate_posterior} of Algorithm \ref{alg:unperturbed} aim to refine \textcolor{black}{the training data by collecting simulated parameters and simulated data that are conditional to the} observed $\by_o^{(i)}$. This is done by sampling \textcolor{black}{$N/M$} times from the learned amortized posterior,  by first inputing $\by^{(i)}\leftarrow \by^{(i)}_o$ inside $q_{\bphi_0}(\bc, \bkappa, \bxi \g \by^{(i)})$, and then sampling $(\bc_j^{(i)}, \bkappa_j^{(i)}, \bxi_j^{(i)}) \sim q_{\bphi_0}(\bc, \bkappa, \bxi \g \by_o^{(i)})$  (step \ref{alg:sample_surrogate_posterior}). At this stage, sampling from $q_{\bphi_0}(\bc, \bkappa, \bxi \g \by_o^{(i)})$  is trivial to do (the surrogate posterior is a Gaussian mixture model), and then, conditionally on each of the \textcolor{black}{$N/M$} posterior draws, in step \ref{alg:simulate_surrogate_posterior} we produce a corresponding simulated observation $\by_j^{(i)} \sim p(\by \g \bc_j^{(i)}, \bkappa_j^{(i)},\bxi_j^{(i)})$  (implicitly from the computer simulator for \eqref{eq:obs_model}). Note that we use the notation $\bkappa_j^{(i)}$ and $\bxi_j^{(i)}$ to emphasize that the shared parameters $\bkappa$ and $\bxi$ are sampled from the surrogate posterior conditional on $\by_o^{(i)}$. Hence, $\bkappa_j^{(i)}$ and $\bxi_j^{(i)}$ depend only on the observation $\by_o^{(i)}$ and not on all $M$ individuals in the data $\by_o$. The surrogate posterior samples and the simulated observations are then collected into a new training data set $\mathcal{D}_1^{(i)} = \{(\bc_j^{(i)}, \bkappa_j^{(i)}, \bxi_j^{(i)}), \by_j^{(i)}\}_{j=1}^{\textcolor{black}{N/M}}$. This is repeated for each individual $i$, and GLLiM is then trained on the union of training data sets from all individuals $\mathcal{D}_1 = \bigcup_{i=1}^{M} \mathcal{D}_1^{(i)}$, to return new estimators $\tilde{\phi}_1$ and ${\phi}_1$, and hence \textcolor{black}{a new surrogate likelihood $q_{\tilde\bphi_1}(\by \g \bc, \bkappa, \bxi)$ (where any observed individual $\by_o^{(i)}$ can be plugged in place of $\by$)} and a corresponding surrogate posterior $q_{\bphi_1}(\bc, \bkappa, \bxi \g \by)$ (steps \ref{alg:train_likelihood_surrpost}-\ref{alg:train_post_surrpost}). This concludes round $r=1$. \textcolor{black}{Training GLLiM on the full data set $\mathcal{D}_1$ is preferable to training $M$ specialized surrogates, separately, each on a corresponding data set $\mathcal{D}_1^{(i)}$. This is preferable since a single GLLiM model needs to be trained, and this improves scalability with respect to the number of individuals $M$. While fitting a single GLLiM model may not be particularly computationally demanding, instead doing so $M$ times, independently, once for each $\mathcal{D}_1^{(i)}$, could results in a very high computational effort when $M$ is very large. In addition, the single learned GLLiM model benefits from training on a larger dataset from the same underlying model, which could improve the approximation. Individual likelihood and posterior approximations can then be obtained simply by imputing data $\by_o^{(i)}$ and parameters $\bc^{(i)}$ from a specific individual $i$.} Notice that it is possible for samples from the initial surrogate $(\bc_j^{(i)}, \bkappa_j^{(i)}, \bxi_j^{(i)}) \sim q_{\bphi_0}(\bc, \bkappa, \bxi \g \by_o^{(i)})$  to ``leak'' outside of the prior's support, if this is bounded (the leaking is not possible in later steps where a Metropolis-Hastings regularization is implemented), therefore step \ref{alg:sample_surrogate_posterior} in Algorithm \ref{alg:unperturbed} could also be written to include a rejection procedure whenever $\pi(\bc_j^{(i)}, \bkappa_j^{(i)}, \bxi_j^{(i)})=0$, so that the sampling is repeated until $N/M$ parameters having $\pi(\bc_j^{(i)}, \bkappa_j^{(i)}, \bxi_j^{(i)})>0$ are collected.

\textcolor{black}{Next come the remaining rounds $r=2,\dotsc,R$}, \textcolor{black}{and unlike for $r=1$, posterior sampling now becomes less immediate, since the product of the surrogate likelihood with the prior is not necessarily  proportional to the density of a Gaussian mixture. This is why we now incorporate a Metropolis-within-Gibbs strategy}. After initializing $\bc, \bkappa, \bxi$ and $\bseta$ (step \ref{alg:init_gibbs}), the remaining part of the algorithm concerns obtaining further refined surrogate likelihood and posterior, where training data includes $N_g$ samples from the Gibbs steps in eq. \eqref{eq:gibbs_gllim_c}-\eqref{eq:gibbs_gllim_eta}, by utilizing the surrogate likelihood $q_{\tbphi_1}(\by \g \bc, \bkappa, \bxi)$ and posterior  $q_{\bphi_1}(\bc, \bkappa, \bxi \g \by)$. First, the individual parameters $\bci$ in \eqref{eq:gibbs_gllim_c} are sampled for subject $i$ independently of other subjects  (step \ref{alg:gibbs_c}) according to an independence-Metropolis-Hastings algorithm \citep{robert2004monte}, which is detailed in Algorithm \ref{alg:MH_c_unperturbed} and that we justify further below. Note that the surrogate posterior $q_{\bphi}(\bc \g \by_o^{(i)})$, where the components corresponding to $\bkappa$ and $\bxi$ have been \textcolor{black}{marginalized out}, is being used as a \textit{self-tuned} proposal distribution in Algorithm \ref{alg:MH_c_unperturbed} (it is ``self-tuned'' because (i) the means and covariance matrices of its components have been automatically provided by the EM procedure, and (ii) it is conditional to the observed data $\by_o^{(i)}$). 

Proposing independently from $q_{\bphi}(\cdot \g \by_o^{(i)})$ is a key feature of SeMPLE: since this proposal function is a mixture model, it is particularly suited for the exploration of multimodal posteriors, and the fact that it has been derived from the same training data as for the surrogate likelihood (and has the same number of components as the mixture model of the likelihood) makes it an appropriate proposal sampler. Moreover, $q_{\bphi}(\cdot \g \by_o^{(i)})$ is conditional on the individual-specific $\by_o^{(i)}$, and this makes it particularly well-informed for the task of sampling specifically $\bc^{(i)}$ in \eqref{eq:gibbs_gllim_c}. Next, in step \ref{alg:gibbs_kappa_xi}, $\bkappa$ and $\bxi$ are updated by targeting the posterior proportional to $\pi(\bkappa_{j-1}, \bxi_{j-1}) \prod_{i=1}^{M} q_{\tilde\bphi_{r-1}}(\by_o^{(i)} \g \bc_{j}^{(i)}, \bkappa_{j-1}, \bxi_{j-1})$, however here the sampling is instead carried out using NUTS, and we justify this specific choice further below. To run NUTS we use \texttt{Stan} \citep{Stan} through the interfaces \texttt{RStan} \citep{RStan} and \texttt{cmdstanr} \citep{cmdstanr}. The final step of the Gibbs sampler (step \ref{alg:gibbs_eta} in Algorithm \ref{alg:unperturbed}) does not involve surrogate likelihoods, and can therefore be easily dealt via NUTS, or can be sampled exactly by using  Normal-Gamma conjugate priors, as detailed in Appendix \ref{sec:conjugate_normal_gamma}. We plug the newly obtained samples into the computer simulator for \eqref{eq:obs_model} to obtain $\by_j^{(i)} \sim p(\by \g \bc_j^{(i)}, \bkappa_j, \bxi_j), i=1,...,M$. The procedure is then repeated to obtain $N_g$ samples of each parameter and corresponding simulated observations. The Gibbs samples and corresponding simulated observations for all $M$ individuals are collected into a data set $\mathcal{D}_r = \Big\{\{\bc_j^{(i)}\}_{i=1}^M, \bkappa_j, \bxi_j, \{\by_j^{(i)}\}_{i=1}^M\Big\}_{j=1}^{N_g}$, and GLLiM is then trained on all data sets up until this point $\bigcup_{\tilde r=1}^{r} \mathcal{D}_{\tilde r}$ (with the exception of the prior-predictive data $\mathcal{D}_0$ which is deemed too uninformative to refine surrogates for specific observed data), to obtain an updated surrogate likelihood and posterior  $q_{\tbphi_r}(\by \g \bc, \bkappa, \bxi)$ and $q_{\bphi_r}(\bc, \bkappa, \bxi \g \by)$ respectively (steps \ref{alg:train_likelihood_gibbs}-\ref{alg:train_post_gibbs}). The full Gibbs procedure is then repeated with the new surrogate likelihood and posterior until the final round $R$ is reached. 

The reason for using NUTS in step \ref{alg:gibbs_kappa_xi}, instead of using the surrogate posterior $q_{\bphi}(\bc^{(i)}, \bkappa, \bxi \g \by_o^{(i)})$ as independence proposal distribution, is that when new values for $\bkappa$ and $\bxi$ are sampled, they need to be proposed conditional on all observations in $\by_o=(\by_o^{(1)}, \dotsc, \by_o^{(M)})$. This is not compatible with the individual surrogate posterior $q_{\bphi}(\bc^{(i)}, \bkappa, \bxi \g \by_o^{(i)})$ that we learn with SeMPLE, as each individual surrogate distribution depends only on the observation of one individual $\by_o^{(i)}$. To circumvent this, one would need to train an additional ``global'' surrogate model $q_{\bphi}(\bc,\bkappa, \bxi \g \by_o)$ that is conditional on the entire observed data set of $M$ subjects. However, for increasing $M$, the dimension of $\by_o$ may become too large to fit a GLLiM model to ``raw data'' (ie non-summarized data) in this way. Hence, we use NUTS to propose new values for $\bkappa$ and $\bxi$. This problem may be relaxed if the inference was based not on the raw observed data, but on summary statistics thereof, and hence reduce the data-dimensionality problem as commonly done in approximate Bayesian computation literature. In Section \ref{sec:scalable-semple} we illustrate an opportunity to obtain a much more scalable Gibbs sampler, that corresponds to a slightly different model formulation.

\textcolor{black}{Notice that the number of components $K$ may progressively reduce during several rounds of Algorithm \ref{alg:unperturbed}, as the implementation of the EM step in the \texttt{xLLiM} package automatically removes irrelevant mixture components having probability $\tilde{\omega}_k=0$ (within floating point accuracy).}

\subsection{A scalable SeMPLE approach for mixed-effects models}\label{sec:scalable-semple}

The approach we have described thus far offers a compelling framework for Bayesian inference in mixed-effects models by automatically constructing surrogate likelihoods and posteriors, including MCMC proposal samplers, that are both efficient to evaluate and fast to sample from. However, it is possible to obtain considerable gains in terms of computational scalability, by slightly reducing the generality of the  mixed-effects model, namely not assume any shared parameters $(\bkappa,\bxi)$, and instead allow these to become part of the individual parameters $\bc^{(i)}$, and hence be random effects. This means that the second Gibbs step can be avoided.
This approach is similar to the ``perturbed'' model in \cite{persson2022}; however, while they fixed the variance of certain parameters, we infer the population variance for all parameters. The three-steps Gibbs sampler  \eqref{eq:gibbs_gllim_c}-\eqref{eq:gibbs_gllim_eta} is replaced by the following two-steps Gibbs sampler
\begin{flalign}
    &\text{step 1: \,}\hat{\pi}(\bc^{(i)} \g \bseta, \by^{(i)})  \propto\pi(\bc^{(i)} \g \bseta) \, q_{\tilde{\phi}}(\by^{(i)} \g \bc^{(i)}),&& \label{eq:gibbs_gllim_c_only_individual}\\
    & \qquad \qquad\qquad\qquad \qquad\qquad i=1,...,M. \nonumber \\
    &\text{step 2: \,}\pi(\bseta \g \bc, \by)  \propto \pi(\bseta) \prod_{i=1}^{M} \pi(\bc^{(i)} \g \bseta).&&\label{eq:gibbs_gllim_eta_only_individual}
\end{flalign}
This ``reduced'' Gibbs sampler, where $\bc^{(i)}=(\cdots,\bkappa^{(i)},\bxi^{(i)})$, has the potential to scale much better with the number of individuals in the data set. The reason why it is beneficial for scaling to remove the fixed-effects is two-fold. First, this removes the need to target the full likelihood $\prod_{i=1}^{M}q_{\tilde\bphi}(\by^{(i)} \g \bc^{(i)}, \bkappa, \bxi)$, which is the product of all individual likelihoods. In fact, when the number of individuals $M$ grows, the automatic differentiation tool needed to perform NUTS will have to take care of differentiating with respect to the full likelihood, which is going to be a very long expression, where the many algebraic operations for its computation involve large matrices such as $\tilde{\boldsymbol A}_k$, $\tilde{\boldsymbol \Sigma}_k$ etc. Evaluating the gradient of such a large expression at every proposed parameter, will cause a considerable computational overhead. Of course, the complexity would be greatly reduced if the inference was based on data-summarization $S(\by)$ rather than $\by$, see the end of Section \ref{sec:background}, since the dimensions of eg. $\tilde{\boldsymbol A}_k$ and $\tilde{\boldsymbol \Sigma}_k$ depend on the dimensions of $\by$. Secondly, when using NUTS to propose fixed-effects $\bkappa$ and $\bxi$ in \eqref{eq:gibbs_gllim_kappaxi}, it is not possible to utilize one of the main benefits of our methodology, which is the self-tuning proposal sampler (and surrogate posterior) $q_{\bphi}$, that has desirable abilities to explore multimodal surfaces \citep{haggstrom2024}. On the other hand, when considering all parameters to be individual as in \eqref{eq:gibbs_gllim_c_only_individual}-\eqref{eq:gibbs_gllim_eta_only_individual}, the proposal sampler $q_{\bphi}$ can be used to propose all the $\bc^{(i)}$ (which include $(\bkappa^{(i)},\bxi^{(i)})$). 

\section{Examples on simulated and real biological data}

We evaluate the performance of SeMPLE for mixed-effects models through \textcolor{black}{three case studies based on two models}. First, we examine a state-space SDEMEM with latent dynamics driven by Ornstein-Uhlenbeck SDEs. This example is particularly relevant, as exact Bayesian inference is feasible without relying on pseudomarginal methods, providing a ``gold-standard'' reference posterior for comparison with SeMPLE inference. Next, we present two case studies where the reference posterior is obtained using a pseudomarginal (particle MCMC) sampler. These examples utilize the SDE model from \cite{pieschner2022identifiability}, which describes translation kinetics following mRNA transfection. One case involves simulated data, and in another case we use real data from \cite{frohlich2018multi}. The model is two-dimensional, with only one observed component, and SeMPLE inference is compared to exact Bayesian (pseudomarginal) inference obtained using the PEPSDI framework (\cite{persson2022}). \textcolor{black}{PEPSDI (\texttt{P}articles \texttt{E}ngine for \texttt{P}opulation \texttt{S}tochastic \texttt{D}ynam\texttt{I}cs) provides a useful benchmark as, to the best of our knowledge, PEPSDI is the only software implementing pseudomarginal schemes for state-space SDEMEMs, including diagnostics for determining an appropriate number of particles, and several proposal kernels for pMCMC.}

\subsection{Ornstein-Uhlenbeck state-space model}
The Ornstein-Uhlenbeck process is defined by the following SDE, 
\begin{equation} 
    \label{eq:ou_model}
    d X^{(i)}_{t} = c_1^{(i)}  (c_2^{(i)} - X_t^{(i)}) dt + c_3^{(i)} dW_t^{(i)}, 
\end{equation}
where the $\{W_t^{(i)}\}_{t\geq 0}$ are independent Wiener processes. In this example, $\{X_t^{(i)}\}_{t\geq 0}$ is one-dimensional. We consider the following state-space model, for $i = 1,...,M$:
\begin{align} \label{eq:ou_model_noise}
    \left\{
        \begin{array}{ll}
        d X_{t}^{(i)} &= c_1^{(i)}  (c^{(i)}_ 2 - X_t^{(i)}) dt + c_3^{(i)} dW_t^{(i)} \\
        Y_t^{(i)} &= X_t^{(i)} + \epsilon_t^{(i)}, \quad \epsilon_t^{(i)} \sim \mathcal{N}(0, \xi^2), 
        \end{array}
    \right.
\end{align}
where $\{Y_t\}_{t\geq 0}$ is the observed process. The transition densities for the latent dynamics are known, and hence the Euler-Maruyama discretization is not needed to simulate the SDE \eqref{eq:ou_model} numerically. Instead we use the following exact simulation scheme, which is induced by the exact transition density
\begin{align}
    X^{(i)}_{t+\Delta_t} &= c^{(i)}_2 + (X^{(i)}_t - c^{(i)}_2)  e^{-c^{(i)}_1 \Delta_t} + \nonumber \\& c^{(i)}_3\sqrt{\frac{(1-e^{-2c^{(i)}_1 \Delta_t})}{2c^{(i)}_1} } \times u^{(i)}_t,
\end{align}
where $u^{(i)}_t \overset{\mathrm{iid}}{\sim} \mathcal{N}(0,1)$.

\subsubsection{Inference setup} \label{sec:ou_setup}
We consider an inference setting similar to the one in \cite{wiqvist2021} and \cite{persson2022}. Data were simulated for $M=40$ individuals at 50 equidistant time points from $t=0.2$ to $t=10$ ($\Delta_t = 0.2$), and with initial value $X_0 = 0$ at $t=0$. We set a Gaussian population distribution $\log(c_1^{(i)}, c_2^{(i)}, c_3^{(i)}) \sim \mathcal{N}(\bmu, \btau^{-1})$, where the true data generating values for the population parameters were set to $\bmu = (-0.7, 2.3, -0.9)$ and $\btau = (4,10,4)$. Similarly to \cite{wiqvist2021}, the prior of $\bseta = (\mu_1, \mu_2, \mu_3, \tau_1, \tau_2, \tau_3)$ was set to be $\pi(\bseta)=\prod_{j=1}^3\pi(\mu_j|\tau_j)\pi(\tau_j)$, where the $\pi(\mu_j|\tau_j)$ and the $\pi(\tau_j)$ are in equation \eqref{eq:ou_normal_gamma}:
\begin{align}
    \begin{cases}
    \mu_j | \tau_j \sim \mathcal{N}(\mu_{0_j}, (\lambda_j \tau_j)^{-1}), \qquad j=1,2,3, \label{eq:ou_normal_gamma}\\ 
    \tau_j \sim Ga(\alpha_j, \beta_j), 
    \end{cases}
\end{align}
\textcolor{black}{where the Gamma distribution is parameterized by the shape $\alpha_j$ and rate $\beta_j$.}
The prior parameter values can be found in Supplementary Material Table \ref{tab:ou_prior_parameters}. The ``Normal-Gamma'' prior \eqref{eq:ou_normal_gamma} allows us to benefit from conjugacy, when sampling $\bseta$ directly from a Normal-Gamma distribution in the third Gibbs sampler step. The prior of $\btau$ is shifted a bit from the setup in \cite{wiqvist2021}, to avoid having small precisions resulting in a large variance in $\bmu$, and consequentially unreasonable prior-predictive simulated data. 
The data-generating value of the noise parameter is  $\log(\xi) = -1.2$, and we used as prior $\log(\xi) \sim \mathcal{N}(0, 1)$. 

\subsubsection{Settings for SeMPLE} \label{sec:ou_semple_settings}
We first determined an appropriate number of mixture components $K$, using the BIC criterion (Appendix \ref{sec:bic}), see Figure \ref{fig:bic_ou-mrna}.
Consequently, we set the starting number of components to be $K=10$. The covariance matrices $\boldsymbol{\Sigma}_k$ and $\tilde{\boldsymbol{\Sigma}}_k$ in the GLLiM models were set to be unconstrained, i.e. fully parameterized.  \textcolor{black}{The number of prior predictive samples is the same as the number of Gibbs samples, i.e. we set $N=N_g=50,000$, and the number of SeMPLE round is $R=4$.}

\subsubsection{Results}

\begin{figure}[ht]
    \centering
    \includegraphics[width=\linewidth]{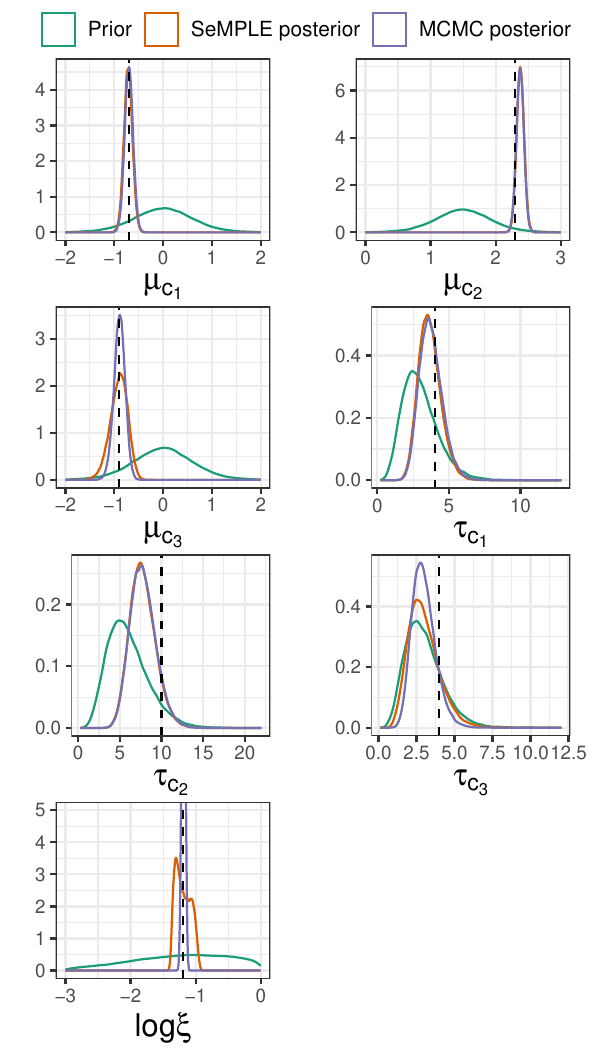}
    \caption{Ornstein-Uhlenbeck: marginal posteriors from 10k posterior samples from MCMC using the exact likelihood (purple) and round $r=4$ of SeMPLE (orange). Priors are in green. The dashed lines are the true parameter values.}
\label{fig:ou_kde_kalman}
\end{figure}

\begin{figure}[ht]
    \centering
    \includegraphics[width=\linewidth]{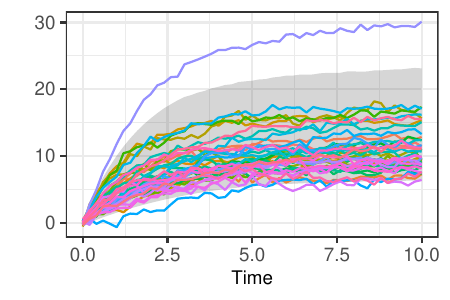}
    \caption{Ornstein-Uhlenbeck: Posterior-predictive simulations from SeMPLE ($r=4$) and data (colored lines, 40 individuals). In grey is the area between the 2.5th and 97.5th percentile from 10k posterior-predictive simulations obtained from SeMPLE. }
    \label{fig:ou_ppc_all_individuals}
\end{figure}

The prior and marginal posteriors for the ``common parameters'' (i.e. shared between subjects) $\bseta$ and $\bxi$ (Figure \ref{fig:ou_kde_kalman}) are obtained using both SeMPLE and exact Bayesian inference (MCMC), where the latter does not employ particle methods. Exact inference is obtained using the three steps of the Gibbs sampler, where the exact likelihood provided by the Kalman filter is used in the Metropolis-within-Gibbs steps \eqref{eq:gibbs_c}-\eqref{eq:gibbs_kappaxi}. Therefore, exact inference is used as a reference to asses the accuracy of the SeMPLE approach in Algorithm \ref{alg:unperturbed}. For exact inference, 200k posterior samples were produced in total, \textcolor{black}{where the first 150k were discarded as burn-in samples, and the remaining 50k samples were used for inference. Inference results obtained with SeMPLE (Figure \ref{fig:ou_kde_kalman}) also used 50k posterior samples. Figure \ref{fig:ou_kde_kalman} reports results from both methods, and we conclude that inference obtained with SeMPLE is overall very satisfying, as even where differences from exact inference appear visually more marked, see the posterior for $\log\boldsymbol{\xi}$, in practical terms these differences are tiny, considering the magnitude of the observations in Figure \ref{fig:ou_ppc_all_individuals}. Moreover, posterior predictive checks in Figure \ref{fig:ou_ppc_all_individuals} confirm the quality of the inference.} Extra results are in Supplementary Material section \ref{sec:app_ou_results}: there, traceplots of the SeMPLE posterior samples show the excellent mixing of the chains from the last round of SeMPLE. Moreover, inference at the individual's level is also excellent, namely, SeMPLE  captures the true values of the data-generating $\bc^{(i)}$'s. 
\textcolor{black}{We have also experimented with different numbers of individuals $M$ ($M=20$ in Figure \ref{fig:OU_M20} and $M=100$ in Figure \ref{fig:OU_M100}): as expected, the marginal posteriors contract around the true data generating parameters, and in particular, for $M=100$, we do learn more about the population precisions $\btau$.}

\subsection{mRNA transfection model} \label{sec:mrna}

We consider the SDE model in a simulation study originally used to describe translation kinetics following mRNA transfection (\cite{pieschner2022identifiability}). The dataset consists of time-lapse microscopy images capturing the fluorescence intensity of cells over at least 30 hours, with measurements taken every 10 minutes (\cite{frohlich2018multi}). During the first hour, cells were incubated with mRNA lipoplexes before being washed to prevent further uptake. As the released mRNA is translated into a green fluorescent protein (GFP), the cell fluorescence is tracked over time.
The SDE below is used to study the translation kinetics of one cell, based on the observed fluorescence trajectory of individual cells after transfection with mRNA encoding for GFP, where the two-dimensional stochastic process $(m(t),p(t))$ where $m(t)$ represents the amount of mRNA molecules at time $t$, and $p(t)$ is the amount of GFP molecules at time $t$.
The SDEMEM is given by
\begin{align}
    \label{eq:mrna_sdemem}
    &d
    \begin{pmatrix}
        m^{(i)} \\
        p^{(i)}
    \end{pmatrix}
    (t)
    = 
    \begin{pmatrix}
        -\delta^{(i)} \cdot m(t)^{(i)} \nonumber \\
        k^{(i)} \cdot m^{(i)}(t) - \gamma^{(i)} \cdot p^{(i)}(t)
    \end{pmatrix}
    dt + \\
    &
    \begin{pmatrix}
        \sqrt{\delta^{(i)} \cdot m^{(i)}(t)} & 0 \\
        0 & \sqrt{k^{(i)} \cdot m^{(i)}(t) + \gamma^{(i)} \cdot p^{(i)}(t)}
    \end{pmatrix}
    d B_t^{(i)} \\
    &(\delta^{(i)}, \gamma^{(i)}, k^{(i)})  \sim \pi(\delta, \gamma, k \g \bseta), \quad i = 1,...,M, \nonumber
\end{align}
where the $B_t^{(i)}$ are two-dimensional standard Brownian motions. It is assumed that all mRNA molecules (within one cell) are released at once from the lipoplexes and denote this initial time point by $t_0$. Before $t_0$, there are neither mRNA nor GFP molecules, and at $t_0$, an amount $m_0$ of mRNA molecules is released, i.e
\[
\begin{pmatrix}
        m^{(i)}(t) \\
        p^{(i)}(t)
    \end{pmatrix}
    = 
    \begin{pmatrix}
        0 \\
        0
    \end{pmatrix}\qquad
    \text{for $t<t_0$}
\]
while $m^{(i)}(t_0)=m_0$ and $p^{(i)}(0)=0$.
We take as observable mapping 
\begin{equation} \label{eq:mrna_obs_model}
    {y}^{(i)}(t_j) = \log(\textrm{scale} \cdot p^{(i)}(t_j) + \textrm{offset})+\varepsilon^{(i)}(t_j),
\end{equation}
where $\varepsilon^{(i)}(t_j)$ is iid Gaussian measurement error  $\varepsilon^{(i)}(t_j) \sim \mathcal{N}(0, \sigma^2)$, $j=1,...,n$, and $t_1,...,t_n$ are observational time instants (see further below). Note that the observations depend only indirectly on process $\{m^{(i)}(t)\}$, and only $\{p^{(i)}(t)\}$ is observed.
The model parameters $(\delta, \gamma, k)$ are treated as random effects parameters that vary between individuals, i.e. $\bc^{(i)} = (\delta^{(i)}, \gamma^{(i)}, k^{(i)})$. The remaining ones, $(m_0, \textrm{scale}, \textrm{offset}, \sigma)$, are treated as fixed but unknown parameters, and we set $\bkappa = (m_0, \textrm{scale}, \textrm{offset})$ and $\bxi = \sigma$. The choice of individual parameters and individual-constant parameters is similar to \cite{arruda2024}. In Section \ref{sec:only-random-effects} we also consider different assumptions where all parameters are set to be random effects, and discuss implications in terms of scalability.

\subsubsection{Inference setup for simulated data} \label{sec:mrna_setup}
In this example, we fix $t_0 = 0$ and do not infer it. This choice enables a direct comparison with the pseudomarginal (particle MCMC) inference produced by the PEPSDI framework \citep{persson2022}, which does not currently support inference of the initial simulation time $t_0$. However, in the real-data case-study in next sections, we infer also $t_0$ (only with SeMPLE). Inference is performed on the log-scale and we assume $\log(\bc^{(i)}) \sim \mathcal{N}(\bmu, \btau^{-1})$, where the true values used to generate data were set to $\bmu = (-0.694, -3, 0.027)$ and $\btau = (10,10,10)$.
It has been shown that SDE modelling improves identifiability of parameters, compared to using a corresponding ODE model \citep{pieschner2022identifiability}. Nevertheless, some parameters are still unidentifiable (for details see supplementary section A.4.1 in \citealp{pieschner2022identifiability}). 

Here the exact transition densities are unavailable, and solutions to the SDE \eqref{eq:mrna_sdemem} are simulated using an Euler-Maruyama scheme implemented in \texttt{Rcpp} \citep{rcpp} from $t=t_0$ to $t=30$, with step size 0.01.
The observed time series is interpolated from the Euler-Maruyama approximation at $n=60$ equidistant time points from $t_1=0.5$ to $t_n=30$. We simulate $M=40$ individuals according to this setup.
The prior distributions are set to be Normal-Gamma for the random effects and Gaussian on the log-scale for the fixed effects. Prior parameters can be found in Supplementary Material section \ref{sec:app_mrna_results}.

\subsubsection{Settings for SeMPLE} \label{sec:mrna_sim_settings_semple}
\textcolor{black}{The starting number of mixture components for the mixture models was set to \textcolor{black}{$K=7$}, according to the Bayesian information criterion (Figure \ref{fig:bic_ou-mrna}}), with covariance matrices for the mixture components specified to be full and unconstrained. \textcolor{black}{The number of prior-predictive samples and Gibbs samples is $N = N_g = 50,000$ and the number of SeMPLE rounds is $R=4$.}

\subsubsection{Setup for the pseudomarginal method PEPSDI}
To assess the quality of the approximate inference obtained by SeMPLE, we run PEPSDI with the setup described in section \ref{sec:mrna_setup}. To avoid typical initialization problems affecting MCMC, when setting starting parameters far from the bulk of the posterior, we initialize PEPSDI at the same (true) parameter values used to generate the observed data. We produce $50,000$ posterior samples, \textcolor{black}{which is the same number of posterior samples produced with SeMPLE. The number of particles was set to 150 for every individual, to reduce the variance of the likelihood estimations and ensure accurate posterior inference}. The initialization of the proposal covariance matrix for the fixed-effects ($\bkappa, \bxi$) was tuned manually to improve the mixing of the Markov chains of these parameters.

\subsubsection{Results from simulated data} \label{sec:mrna_results_sim_data}
\begin{figure}[htb]
    \centering
    \includegraphics[width=\linewidth]{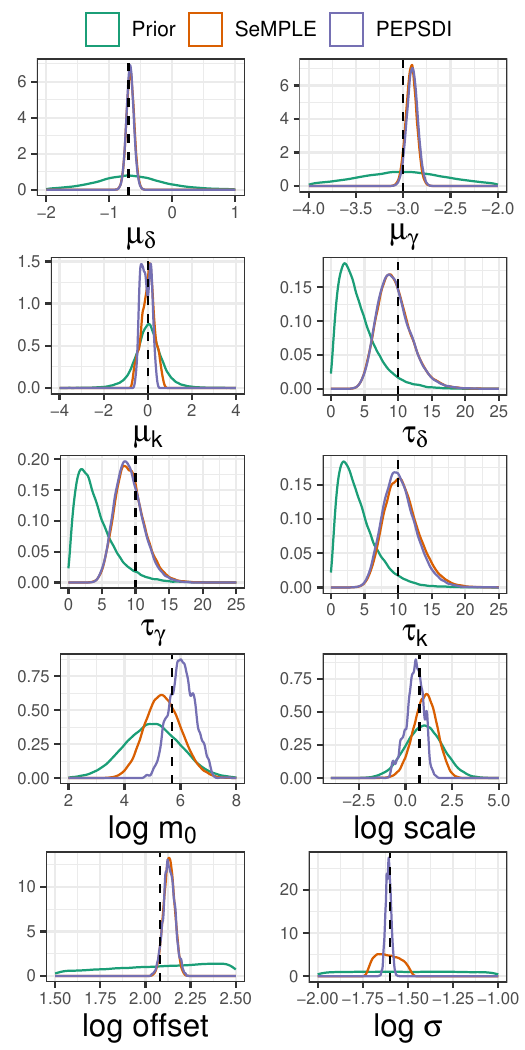}
    \caption{mRNA model with 40 simulated individuals: marginal posteriors obtained with SeMPLE (orange, round $r=4$) and with PEPSDI (purple). Priors are in green. The dashed lines are the true parameter values that were used to generate the observed data.}
    \label{fig:mrna_kde_comp_pepsdi}
\end{figure}

\textcolor{black}{We compare posterior distributions based on 50,000 posterior samples from both PEPSDI and SeMPLE.} The posterior distributions of the population means $\bmu$ and population precisions $\btau$ returned by SeMPLE  are virtually identical to exact (pseudomarginal) inference returned by PEPSDI (Figure \ref{fig:mrna_kde_comp_pepsdi}). The same applies to the $\textrm{offset}$ parameter, \textcolor{black}{and while the marginal posterior for the measurement error's standard deviation $\sigma$ is slightly wider than the PEPSDI posterior, we have noticed that posterior predictive simulations resulting from this are virtually indistinguishable from the observed data (plot not reported).} The posterior plots indicate that the $m_0$ and the ``$\textrm{scale}$'' parameters are more challenging to infer, \textcolor{black}{and this is not specific to SeMPLE, in fact the traceplots from PEPSDI for $m_0$ and $\textrm{scale}$ (Figure \ref{fig:mrna_sim_pepsdi_traceplot_kappaxi}) show difficulties with mixing, and therefore for these two parameters the comparison between SeMPLE and PEPSDI should be taken with a grain of salt. Such difficulty is also evident in the corresponding posterior plots in \cite{arruda2024}.} In addition to the posterior density plots, posterior predictive checks for several exemplary individuals
are given in Supplementary Material. \textcolor{black}{The PEPSDI runtime to produce 50,000 posterior samples was approximately 149 hours}, using the artificially favorable setup where we avoided investing time in the search for a suitable starting value for the parameters.
\textcolor{black}{The SeMPLE runtime throughout the $R=4$ rounds was $77.4$ hours. However, we have verified that already at $r=2$ SeMPLE was producing accurate inference in only $10.4$ hours (see Figure \ref{fig:mrna_kde_comp_pepsdi_r2}).} In addition to this, the determination of the initial $K$ via the BIC took 6 minutes. \textcolor{black}{A comparison between effective sample sizes (ESS, the higher the better)  can be found in Table \ref{tab:ess_iat}. The univariate ESS was computed using the R-package \texttt{LaplacesDemon} \citep{laplacesdemon}, and we also reported the multivariate ESS, via the R-package \texttt{mcmcse} \citep{mcmcse}. Table \ref{tab:ess_iat} shows that, overall, the ESS from SeMPLE are much larger than with pseudomarginal (PEPSDI) inference, and that nearly independent samples are obtained much faster with SeMPLE (around five times faster).} \textcolor{black}{We emphasize that results from a slightly different model can be obtained much more rapidly via SeMPLE, see Section \ref{sec:only-random-effects}, corresponding to the two-steps Gibbs described in Section \ref{sec:scalable-semple}.}

\subsubsection{Inference setup with real data} \label{sec:mrna_setup_real_data}
We now use the same SDEMEM as in the simulated data example to test our method on the real-world dataset from \cite{frohlich2018multi}, to which we refer the reader for details regarding the experimental procedures. This data has also been analysed in \cite{pieschner2022identifiability}, where the initial ODE model was extended to an SDE model with improved parameter identifiability, however in \cite{pieschner2022identifiability} no new inference methodology was presented. We extend their work by applying our novel inference methodology. We consider the first $M=40$ individuals from the data set labeled ``20164027\_mean\_eGFP'' in \cite{frohlich2018multi}. The raw data is then log-transformed in accordance with the observable mapping \eqref{eq:mrna_obs_model}. Note that the real data set has 180 measurements for each individual, as opposed to the 60 measurements in the simulated data setup described in Section \ref{sec:mrna_setup}. To allow more flexibility, the prior distribution of the population parameters are set to be independent Gaussian for the mean $\bmu$ and Gamma-distributed precision $\btau$, instead of a prior Normal-Gamma distribution. The prior parameters can be found in Supplementary Material section \ref{sec:app_mrna_real_data}. Consequentially, it is no longer possible to sample the population parameters $\bseta$ directly, since we cannot exploit conjugacy here. Instead, we use NUTS to sample the population parameters efficiently in the Gibbs sampler. Note that, as opposed to the setup with simulated data, we now infer the initial time point $t_0$ as a parameter that varies between individuals, i.e. $\bc^{(i)} = (\delta^{(i)}, \gamma^{(i)}, k^{(i)}, t_0^{(i)})$; however we can only do so via SeMPLE, as PEPSDI currently does not allow to infer the starting time $t_0$, and therefore we cannot compare results from both methods. This is not a problem per-se, as we have already shown  such a comparison and the SeMPLE reliability in the simulation study.

\subsubsection{Settings for SeMPLE with real data} \label{sec:mrna_real_data_settings_semple}
Similarly to the setting with simulated data in section \ref{sec:mrna_sim_settings_semple}, we set the GLLiM covariance matrices for the mixture components to be full and unconstrained. The number of mixture components was set to \textcolor{black}{$K=9$} according to the BIC (Figure \ref{fig:bic_ou-mrna}). \textcolor{black}{The number of prior-predictive samples was set to $N=50,000$, we produce $N_g=1,000$ posterior Gibbs samples, and the number of SeMPLE rounds is $R=4$}.

\subsubsection{Results from real data example} \label{sec:mrna_results_real_data} 
\begin{figure}
    \centering
    \includegraphics[width=\linewidth]{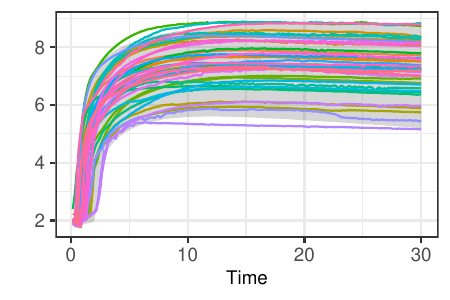}
    \caption{mRNA model with real data: posterior-predictive simulations for 40 individuals using SeMPLE ($r=4$, colored lines are observed data). In grey is the area between the 2.5th and 97.5th percentile from $1,000$ posterior-predictive simulations obtained from SeMPLE.}
    \label{fig:mrna_inference_real_data_all}
\end{figure}

To validate the quality of the inference, we provide posterior predictive simulations for all individuals (Figure \ref{fig:mrna_inference_real_data_all}). The uncertainty about the observed dynamics is well captured and the posterior inference is consistent with the observed data. Additional individual posterior predictive simulation plots can be found in the Supplementary Material section \ref{sec:app_mrna_real_data}.
The SeMPLE runtime  was \textcolor{black}{24 hours}. In addition to this, the runtime to determine the initial $K$ via the BIC for this setup was \textcolor{black}{33 minutes}. Note that the number of posterior samples was reduced, compared to the simulated data setup, to reduce the runtime.

\subsubsection{Scalable approach using exclusively random-effects} \label{sec:only-random-effects}

As explained in Section \ref{sec:scalable-semple}, a major computational bottleneck in the framework presented so far is the second step of the Gibbs sampler \eqref{eq:gibbs_gllim_kappaxi}, where  the full likelihood (the product of all individual likelihoods) is used within NUTS. Here we consider inference with the two-steps ``reduced'' Gibbs sampler from Section \ref{sec:scalable-semple}. 
The population distribution for the translation kinetics model after mRNA transfection can then be rewritten as 
\begin{align}
    &\log(\delta^{(i)}, \gamma^{(i)}, k^{(i)}, t_0\ind, m_0\ind, \textrm{scale}\ind, \textrm{offset}\ind, \sigma\ind) \nonumber \sim \\ 
    &\mathcal{N}(\bmu, \btau^{-1}), \qquad i = 1,...,M.
\end{align}
\textcolor{black}{Therefore, compared to previous analyses, here $(m_0\ind, \textrm{scale}\ind, \textrm{offset}\ind,\sigma\ind)$ are also random effects. }
We run SeMPLE with the same $M=40$ individuals from the real data set as in Section \ref{sec:mrna_results_real_data}. The runtime to produce \textcolor{black}{$1,000$ posterior samples is $1.3$ hours, a significant reduction compared to the corresponding runtime with fixed-effects of 24 hours in section \ref{sec:mrna_results_real_data}. For the purpose of displaying inference results, Figure \ref{fig:mrna_kde_only_individual} shows marginal posterior distributions based on $10,000$ posterior samples. }
We note that for many of the population precisions $\btau$, the posterior distributions are similar to the priors, except for the population precisions $\tau_{t_0}$ and $\tau_{\textrm{offset}}$. This suggests that a data set of 40 individuals is not large enough to infer the population variance of all the model parameters, \textcolor{black}{which we confirm in the next section, where we use larger values for $M$}.

\subsubsection{Runtime scaling} \label{sec:runtime_scaling}
\begin{figure}
    \centering
    \includegraphics[width=\linewidth]{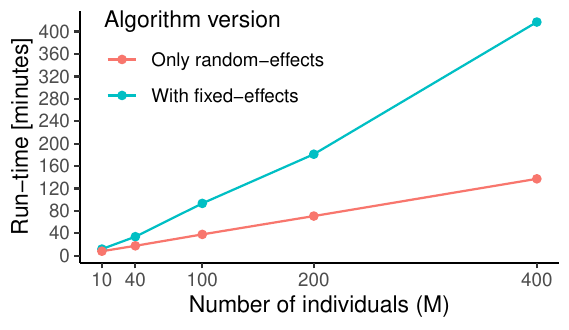}
    \caption{
    SeMPLE runtime to obtain $1,000$ posterior samples as a function of the number of individuals. The blue line corresponds to the algorithm version that also include constant parameters (fixed-effects) and the red line to the version with only individual parameters (random-effects).
    }
    \label{fig:mrna-runtime}
\end{figure}

\begin{figure}
    \centering
    \includegraphics[width=\linewidth]{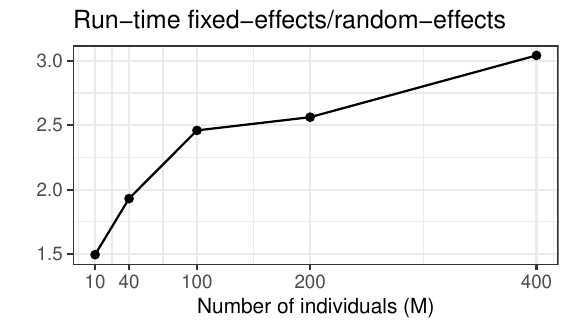}
    \caption{Ratio of SeMPLE runtimes for a model that include fixed-effects and the one having only random-effects, corresponding to the data in Figure \ref{fig:mrna-runtime}.}
    \label{fig:mrna-runtime-ratio}
\end{figure}

To further investigate how the runtime of the SeMPLE algorithm scales with the number of individuals $M$, we perform a simulation study for increasing $M$. We report the run times both with the most general SeMPLE algorithm for the three-steps Gibbs approach (with setup described in Section \ref{sec:mrna_setup_real_data}), and the corresponding setup with the scalable approach described in Section \ref{sec:only-random-effects}. However, to ease the calculations, here we use simulated data with 60 observations for each individual. The number of prior predictive samples is set to $N = 10,000$, and the number of Gaussian mixture components was set to $K = 10$ for both algorithm versions to make comparisons fair, even though previous BIC results (Figure \ref{fig:bic_ou-mrna}) suggests a smaller $K$ could be used. 
We measure the wall-clock runtime to obtain $1,000$ posterior samples from the corresponding Gibbs sampler, and this includes the training of the surrogate likelihood both on prior-predictive data and on samples from the surrogate posterior (Figure \ref{fig:mrna-runtime}). Here, the ``scalable approach'' two-steps Gibbs is denoted with ``only random-effects'', and the three-steps Gibbs with ``with fixed-effects''. For the scalable approach the runtime clearly scales linearly with the number of individuals. This is to be expected since the sampling via Gibbs (i.e. excluding the mixtures fitting)  typically makes up the majority of the runtime, and this scales linearly with $M$ when no parallelization is performed. The runtime of fitting the surrogate models is typically negligible in comparison to the Gibbs sampling. Figure \ref{fig:mrna-runtime-ratio} shows the ratio between the runtimes from Figure \ref{fig:mrna-runtime}, displaying the acceleration achieved when running the two-steps Gibbs instead of the three-steps Gibbs approach.

Connecting to the conjecture that the number of individuals ($M = 40$) in the data set was not large enough to infer the population precision (Figure \ref{fig:mrna_kde_only_individual}), we refer to a simulation study in Supplementary Material using $M=200$  (Figure \ref{fig:mrna_kde_200ind}). With $M=200$ the posterior distributions of the population precisions are now very informative about the location of the true parameter values.

\section{Discussion and conclusion} \label{sec:discussion}
This study introduces a novel simulation-based (Bayesian) inference method for stochastic nonlinear mixed-effects models. More specifically, we focused on mixed-effects driven by SDEs (SDEMEMs). The method addresses the challenge of providing flexible, yet computationally efficient, methodology in this setting. Our SeMPLE methodology builds amortized approximations of the intractable likelihood and of the posterior using Gaussian mixtures of experts, and then proceeds at refining such approximations for given observed data, without using neural conditional estimation, unlike in state-of-the-art SBI methodology. This comes with some advantages, namely easier analytic tractability, the use of well-studied ad-hoc algorithms for their fitting (expectation-maximization), and finally the estimators obtained from a specific conditional density (say the likelihood function), when this is expressed via a Gaussian mixture, they can easily be transferred to other conditional densities (eg the posterior) using closed-form algebraic operations.

For the case studies we considered, including a SDEMEM for translation kinetics model after mRNA transfection, we compared our inference against gold-standard Bayesian inference, either using exact or asymptotically exact (pseudomarginal) MCMC samplers. In all cases we show that with SeMPLE we obtain inference that is very similar to exact Bayes, but with the additional advantage of allowing inference for a much larger number of individuals than could be possible with particle-based MCMC approaches, as further discussed below. In terms of generality, SeMPLE provides full Bayesian inference with the option of treating any model parameter as either a fixed or a random effect, unlike methods where fixed effects are modelled as random effects \textcolor{black}{with zero variance} \citep{arruda2024}. In addition, SeMPLE allows to infer the time point $t_0$ of the discrete jump in the SDE solution as a model parameter, instead of being forced to assume $t_0$ as known, a limitation present in both \cite{pieschner2022identifiability} and \cite{persson2022}.

We have explored the scalability of our methodology with respect to an increasing number of individuals $M$ 
(Section \ref{sec:runtime_scaling}), and we have provided an additional (and slightly less general) version of our method that has an improved scalability 
(Section \ref{sec:only-random-effects}). We have shown that, with the more scalable version of SeMPLE, we can fit SDEMEMs to several hundreds of individuals using a standard laptop, which is particularly notable, \textcolor{black}{and would not have been possible with the pseudomarginal particle MCMC (pMCMC) approaches in, e.g., \cite{persson2022}, without considerable tuning (e.g. tuning for the number of particles, and the usage of ``guided'' paths-solutions for the particle filters)}. Moreover, pMCMC is not an amortized approach and requires reruns for every new considered dataset. Still, for SeMPLE there is room for improvement in terms of the scalability in the number of individuals $M$. For example, fully Bayesian inference for SDEMEMs, with several thousands of individuals, can be computationally demanding \textcolor{black}{for SeMPLE when running on a standard laptop, but could of course be accommodated on a computer cluster}.
\textcolor{black}{
We note that the Gibbs sampler step with the individual parameters in \eqref{eq:gibbs_gllim_c} allows for parallelization, and with a large number of individuals this could reduce the runtime significantly.
}
Another methodology, which targets scalability but using amortized neural density estimators, is in \cite{arruda2024}. However, in \cite{arruda2024} the focus is not on fully Bayesian inference (even though a brief demonstration with Bayesian approaches is given), but instead on maximum likelihood estimation and uncertainty quantification through profile likelihood analysis. It is difficult to construct inference methods for SDEMEMs that are computationally efficient without making simplifications at the cost of generality. In this regard, we identify a need for further research in this field while providing a significant contribution in this direction.

\section*{Acknowledgments}
HH, UP and MC acknowledge support from the Swedish Research Council (Vetenskapsrådet 2019-03924 and 2023-04319). SP and MC acknowledge support from the Swedish Foundation for Strategic Research (FFL15-0238). UP acknowledges support from the Chalmers AI Research Centre (CHAIR). The computations were enabled by resources provided by the National Academic Infrastructure for Supercomputing in Sweden (NAISS), partially funded by the Swedish Research Council through grant agreement no. 2022-06725. 

\section*{Statements and declarations}
\subsection*{Competing interests}
The authors declare no competing interests.

\begin{appendices}
\section{Conjugate priors for the Ornstein-Uhlenbeck model} \label{sec:conjugate_normal_gamma}
For the Ornstein-Uhlenbeck model we use conjugate priors, as a matter of convenience, as it makes it easier to compare the results of SeMPLE with the gold standard inference obtained using the Kalman filter.
With conjugate priors we can sample explicitly from  equation \eqref{eq:gibbs_gllim_eta}, and this is achieved by setting a Normal-Gamma prior distribution on the population parameters $\bseta$, as in the following
\begin{align}
    \begin{cases}
    \mu_j | \tau_j \sim \mathcal{N}(\mu_{0_j}, (\lambda_j \tau_j)^{-1}), \qquad j=1,2,3 \label{eq:normal_gamma_prior_conjugacy}\\ 
    \tau_j \sim Ga(\alpha_j, \beta_j). 
    \end{cases}
\end{align}
We denote the latter prior by Normal-Gamma($\bmu, \boldsymbol\lambda, \boldsymbol\alpha, \boldsymbol\beta$), and we let the population distribution be a Gaussian $\pi(\bc \g \bseta) \sim \mathcal{N}(\bmu, \btau^{-1})$. The distribution that we want to sample from in \eqref{eq:gibbs_gllim_eta} is therefore a Normal-Gamma distribution \citep{murphy_conjugate} with the following parameters 
\begin{align*}
    \textrm{NG}\Big(&\frac{\lambda_j {\mu_0}_j + M \bar \bc_j}{\lambda_j + M}, \lambda_j + M, \alpha_j + M/2, \\ &\beta_j + \frac{1}{2} \sum_{i=1}^M(c^{(i)}_j - \bar \bc_j)^2 + \frac{M \lambda_j}{\lambda_j + M} \frac{(\bar \bc_j - {\mu_0}_j)^2}{2}\Big).
\end{align*}

\section{Determination of \texorpdfstring{$K$}{Lg}}\label{sec:bic}

The number $K$ of components in the Gaussian mixture model  has to be specified prior to performing the EM procedure. A too large value of $K$ may result in overfitting and unnecessary computational effort, while a too small value of $K$ may limit the ability to represent the relationship between $\btheta$ and $\by$. The Bayesian Information Criterion (BIC), used in \cite{deleforge2014} and \cite{haggstrom2024} to guide the selection of  $K$, is given by 
\begin{equation*}
    \label{eq:bic}
    BIC = -2 \mathcal{L}(\hat{\bphi}) + D(\tilde{\bphi}) \log N
\end{equation*}
where $\mathcal{L}(\hat\bphi)$ is the maximized value of the  GLLiM log-likelihood function at the MLE $\hat\bphi$, $D(\Tilde{\bphi})$ is the total number of parameters in the model and $N$ is the number of observations in the training dataset. We wish to select a $K$ returning a small BIC, when GLLiM is fitted with $K$ components to a training dataset 
$\{\btheta_n, \by_n\}_{n=1}^{N}$ obtained by independently sampling parameters $\btheta_n \sim p(\btheta)$ from the prior, and by simulating the corresponding $\by_n \sim p(\by|\btheta_n)$ from the generative model.
When the training data is the set of the $N$ independent $\{\btheta_n, \by_n\}_{n=1}^{N}$, the GLLiM log-likelihood is given by
\begin{equation*}
    \mathcal{L(\tilde{\bphi})} = \sum_{n=1}^{N} \log q_{\tilde{\bphi}}(\by_n, \btheta_n),
\end{equation*}
with
\begin{equation*}
    q_{\tilde{\bphi}}(\by_n, \btheta_n) = \sum_{k=1}^{K} 
    \mathcal{N}(\by_n; \tilde{\boldsymbol A}_k \btheta + \tilde{\boldsymbol b}_k, \tilde{\boldsymbol \Sigma}_k)\;  \mathcal{N}(\btheta_n; \tilde{\boldsymbol{\nu}}_k, \tilde{\boldsymbol \Gamma}_k)  \;
    \pi_k \; .
\end{equation*}
Note that within our framework $\btheta = (\bc, \bkappa, \bxi)$ and thus $l = q+p+s$, where $\btheta \in \mathbb{R}^l$, and $\by \in \mathbb{R}^{d_o \times n}$, where $d_o$ is the dimension of the observation at a specific time point and $n$ is the number of time points in the observation. The number of parameters that GLLiM needs to estimate is
\begin{align} \label{eq:gllim_total_num_param}
    D(\tilde{\bphi}) =& (K-1) + K\big(d_o n (q+p+s) + d_o n + \nonumber \\ & (q+p+s) +\operatorname{nbpar}_\Sigma + \operatorname{nbpar}_\Gamma\big),
\end{align}
where $\operatorname{nbpar}_\Sigma$ and $\operatorname{nbpar}_\Gamma$ are the number of parameters in the covariance matrices $\tilde{\Sigma}_k$ and $\tilde{\Gamma}_k$, respectively. The covariance structure of the matrices $\tilde{\Sigma}_k$ and $\tilde{\Gamma}_k$ can be constrained to reduce the number of parameters that GLLiM needs to estimate.
The \texttt{xLLiM} package \citep{xllim}, that we use to run Expectation-Maximization when fitting GLLiM models, allows the $\tilde{\boldsymbol{\Sigma}}_k$'s to be set as isotropic, diagonal or full matrices, and set all equal or varying with $k$.
The setup we considered for the covariance matrices can be found in the sections pertaining each of the considered examples. 

\begin{figure}[htb]
    \centering
\begin{subfigure}[b]{\linewidth}
    \includegraphics[width=0.8\textwidth]{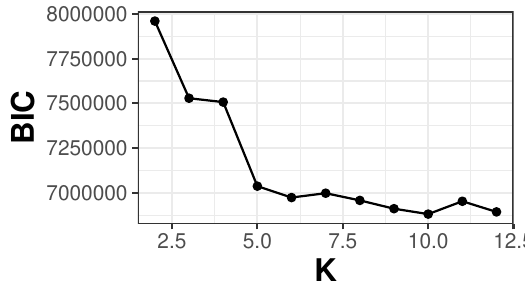}
    \caption{}
\end{subfigure}
\begin{subfigure}[b]{\linewidth}
    \includegraphics[width=0.8\textwidth]{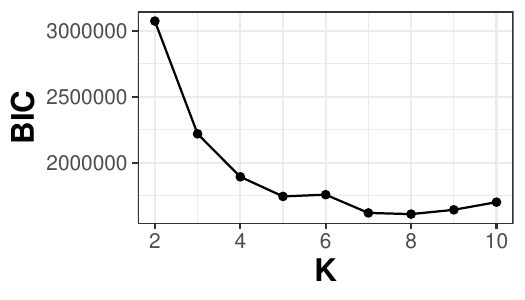}
    \caption{}
\end{subfigure}    
\begin{subfigure}[b]{\linewidth}
    \includegraphics[width=0.8\textwidth]{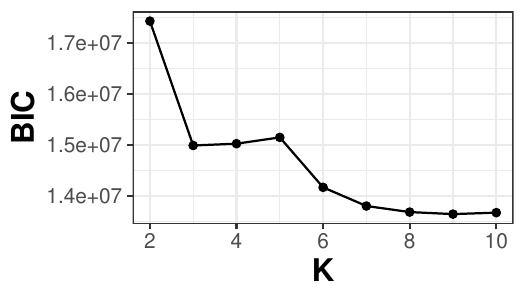}
    \caption{}
\end{subfigure}  
\caption{Bayesian information criterion as a function of $K$. (a) Ornstein-Uhlenbeck model. (b) mRNA model with simulated data. (c) mRNA model with real data (40 subjects).}
\label{fig:bic_ou-mrna}
\end{figure}

\end{appendices}

\bibliography{refs}

\newpage
\clearpage

\newcommand{\beginsupplement}{%
        \setcounter{table}{0}
        \renewcommand{\thetable}{S\arabic{table}}%
        \setcounter{figure}{0}
        \renewcommand{\thefigure}{S\arabic{figure}}%
    }
\makeatletter
\renewcommand \thesection{S\@arabic\c@section}
\renewcommand\thetable{S\@arabic\c@table}
\renewcommand \thefigure{S\@arabic\c@figure}
\makeatother

\onecolumn 
\newcommand{\adjustedfigwidth}{0.5}
\beginsupplement
\section{Supplementary Material}

\subsection{Ornstein-Uhlenbeck} \label{sec:app_ou_results}
\begin{figure}[H]
    \centering
    \begin{subfigure}[b]{0.475\linewidth}
        \centering
        \includegraphics[width=0.5\linewidth]{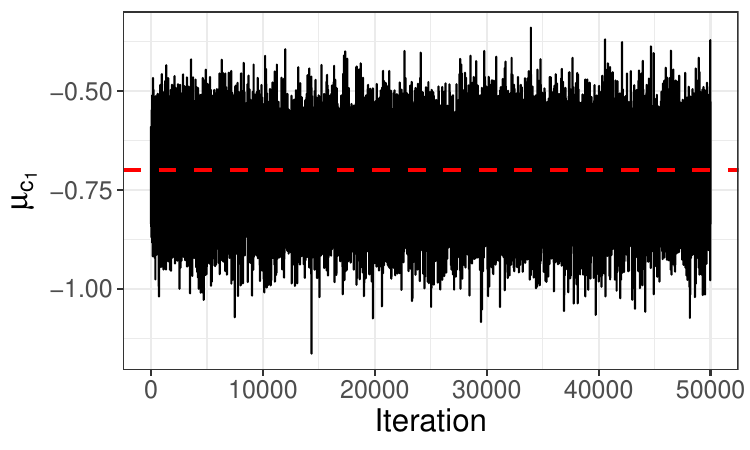}
          \caption{$\mu_{c_1}$}%
        \label{fig:ou_traceplot_mu1}
    \end{subfigure}
    \begin{subfigure}[b]{0.475\linewidth}
        \centering
        \includegraphics[width=0.5\linewidth]{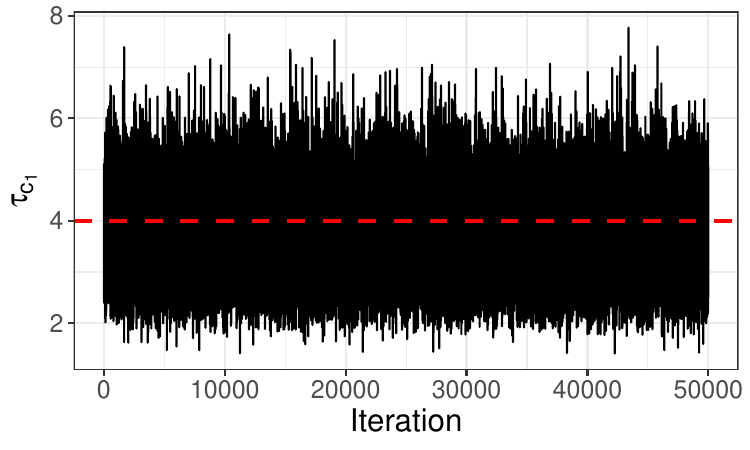}
          \caption{$\tau_{c_1}$}%
        \label{fig:ou_traceplot_tau1}
    \end{subfigure}
    \begin{subfigure}[b]{0.475\linewidth}
        \centering
        \includegraphics[width=0.5\linewidth]{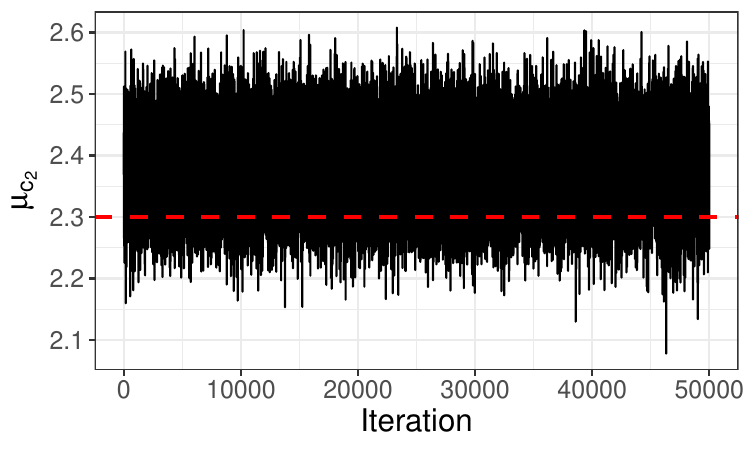}
          \caption{$\mu_{c_2}$}%
        \label{fig:ou_traceplot_mu2}
    \end{subfigure}
    \begin{subfigure}[b]{0.475\linewidth}
        \centering
        \includegraphics[width=0.5\linewidth]{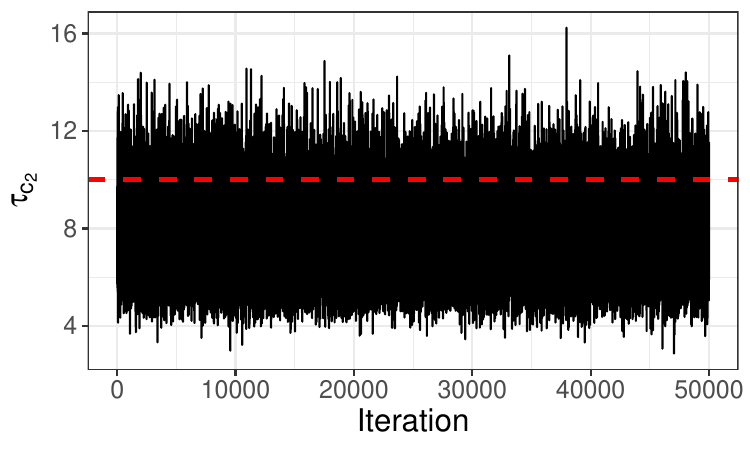}
         \caption{$\tau_{c_2}$}%
        \label{fig:ou_traceplot_tau2}
    \end{subfigure}
    \begin{subfigure}[b]{0.475\linewidth}
        \centering
        \includegraphics[width=0.5\linewidth]{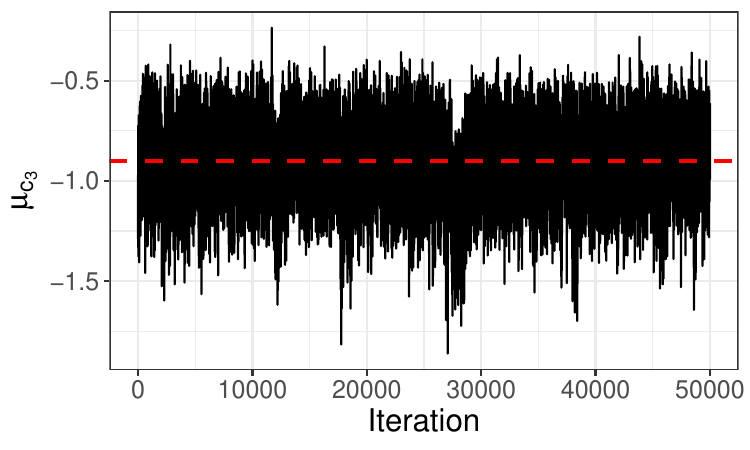}
          \caption{$\mu_{c_3}$}%
        \label{fig:ou_traceplot_mu3}
    \end{subfigure}
    \begin{subfigure}[b]{0.475\linewidth}
        \centering
        \includegraphics[width=0.5\linewidth]{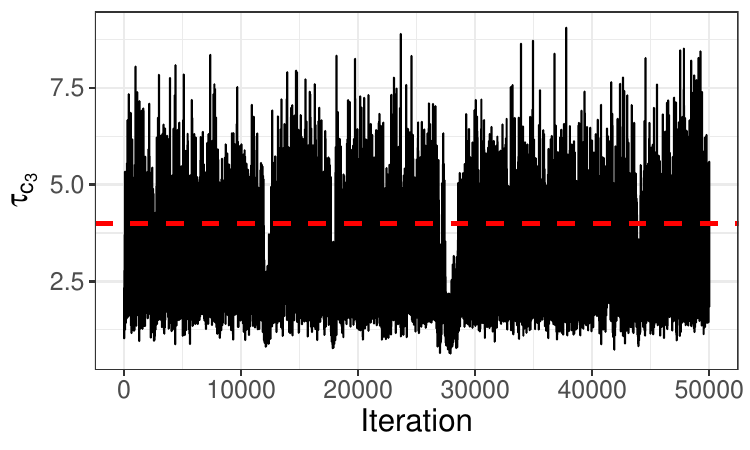}
         \caption{$\tau_{c_3}$}%
        \label{fig:ou_traceplot_tau3}
    \end{subfigure}
    
    \begin{subfigure}[b]{0.475\linewidth}
        \centering
        \includegraphics[width=0.5\linewidth]{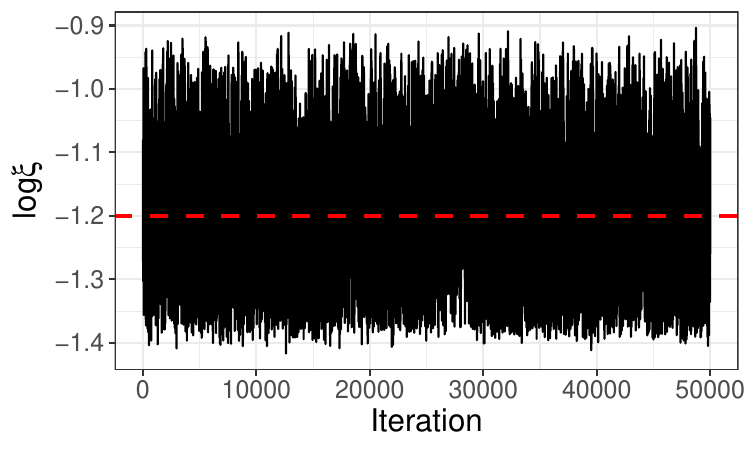}
         \caption{$\log\xi$}%
        \label{fig:ou_traceplot_xi1}
    \end{subfigure}

    \caption{Ornstein-Uhlenbeck, traceplots of posterior samples from round 4 of SeMPLE  for the ``common'' parameters $\bseta = (\mu_1, \mu_2, \mu_3, \tau_1, \tau_2, \tau_3)$ and $\bxi$. The red dashed line shows the true value used to generate the observed data.}
    \label{fig:ou_traceplots}
\end{figure}

The prior parameters can be found in Table \ref{tab:ou_prior_parameters}.
\begin{table}[h]
\begin{tabular}{l|llll}
j & $\mu_{0_j}$ & $\lambda_j$ & $\alpha_j$ & $\beta_j$ \\
\hline
1  &      0       & 1 & 6 & 2 \\
2  &       1.5    & 1 & 6 & 1 \\
3  &         0    & 1 & 6 & 2
\end{tabular}
\caption{Ornstein-Uhlenbeck prior distribution parameters.}
\label{tab:ou_prior_parameters}
\end{table}
Figure \ref{fig:ou_traceplots} shows the traceplots obtained from the last round (round 4) of SeMPLE for the ``common'' (shared between subjects) parameters $\bseta = (\mu_1, \mu_2, \mu_3, \tau_1, \tau_2, \tau_3)$ and $\bxi$.

\begin{figure}[H]
    \centering
    \begin{subfigure}[b]{0.55\linewidth}
        \centering
        \includegraphics[width=\linewidth]{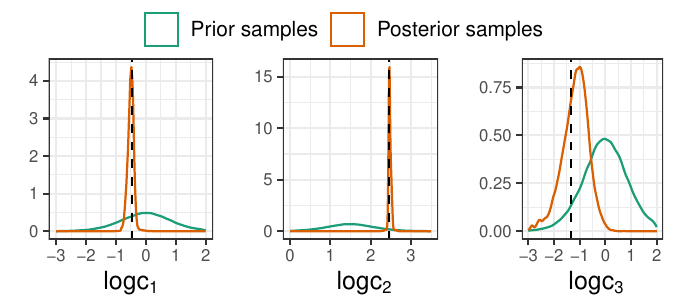}
    \end{subfigure}
    \begin{subfigure}[b]{0.55\linewidth}
        \centering
        \includegraphics[width=\linewidth]{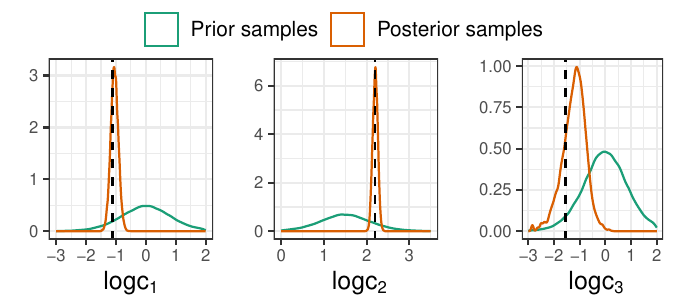}
    \end{subfigure}
    \begin{subfigure}[b]{0.55\linewidth}
        \centering
        \includegraphics[width=\linewidth]{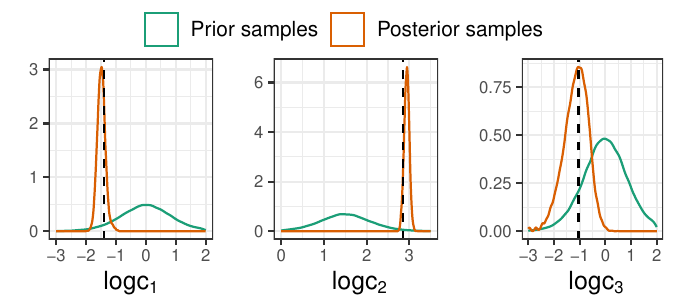}
    \end{subfigure}
    \begin{subfigure}[b]{0.55\linewidth}
        \centering
        \includegraphics[width=\linewidth]{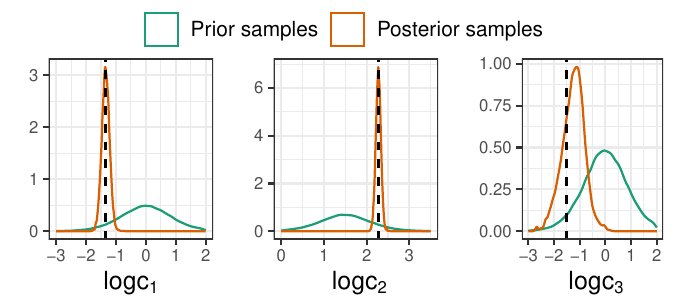}
    \end{subfigure}
    \begin{subfigure}[b]{0.55\linewidth}
        \centering
        \includegraphics[width=\linewidth]{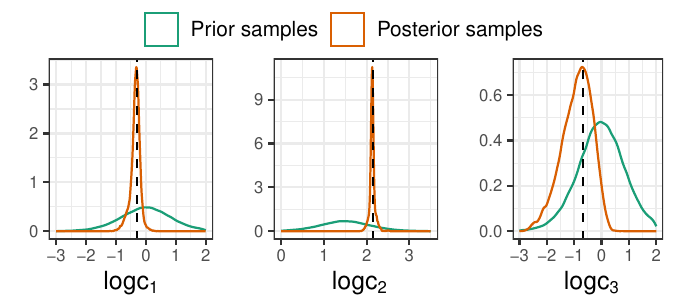}
    \end{subfigure}
    \caption{Ornstein-Uhlenbeck, inference from SeMPLE: from top to bottom we report inferences from individuals 1 to 5. In orange are the kernel density estimates of posterior samples from round $r=4$ of SeMPLE for the individual parameters $\log\bc^{(i)}$. The prior is in green. The red dashed lines show the true parameters for the specific individual.}
    \label{fig:ou_kde_ind}
\end{figure}

Inference for individual parameters $\log\bc^{(i)}=(\log c^{(i)}_1,\log c^{(i)}_2,\log c^{(i)}_3)$ is in Figure \ref{fig:ou_kde_ind}. We considered $M=40$ subjects and therefore, given the large numbers of possible plots to display, as an illustration, we report inference for the first five subjects only, $i=1,...,5$. The inference from SeMPLE is informative about the true values of the $\log \bc^{(i)}$'s, while clearly producing a noticeable learning compared to the corresponding priors.

\begin{figure}[H]
    \centering
    \begin{subfigure}[b]{0.475\linewidth}
        \centering
        \includegraphics[width=0.5\linewidth]{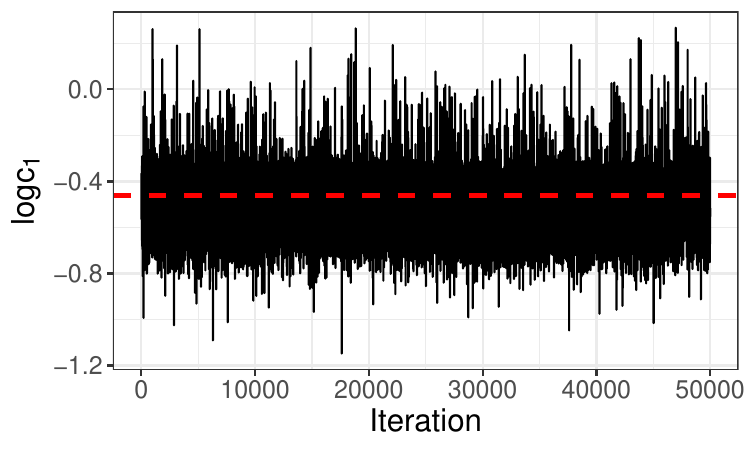}
        \caption{$\log{c_1^{(1)}}$}
    \end{subfigure}
        \begin{subfigure}[b]{0.475\linewidth}
        \centering
        \includegraphics[width=0.5\textwidth]{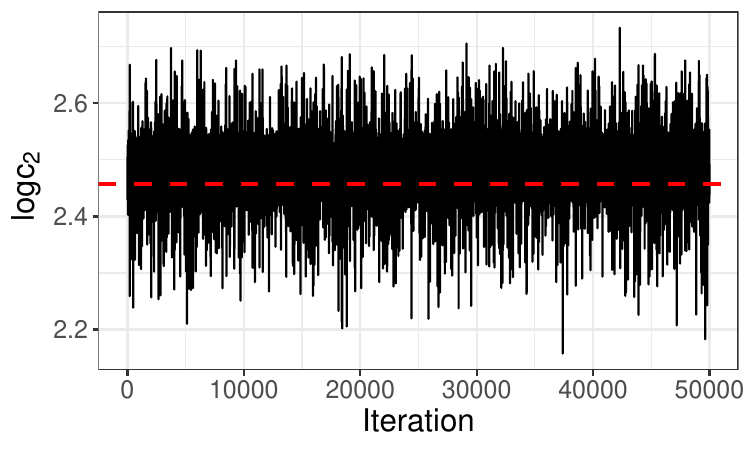}
        \caption{$\log{c_2^{(1)}}$}
    \end{subfigure}
        \begin{subfigure}[b]{0.475\linewidth}
        \centering
        \includegraphics[width=0.5\textwidth]{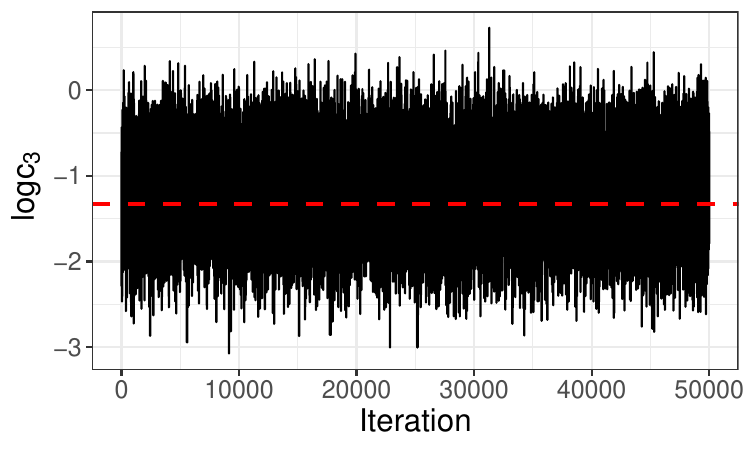}
        \caption{$\log{c_3^{(1)}}$}
    \end{subfigure}
    \caption{Ornstein-Uhlenbeck, traceplots of inference from SeMPLE for individual parameters for individual 1 out of 40: traceplots are from round 4 of SeMPLE. The red dashed line shows the true value of this specific individual.}
    \label{fig:ou_traceplot_c}
\end{figure}

\begin{figure}[H]
    \centering
    \includegraphics[width=0.8\linewidth]{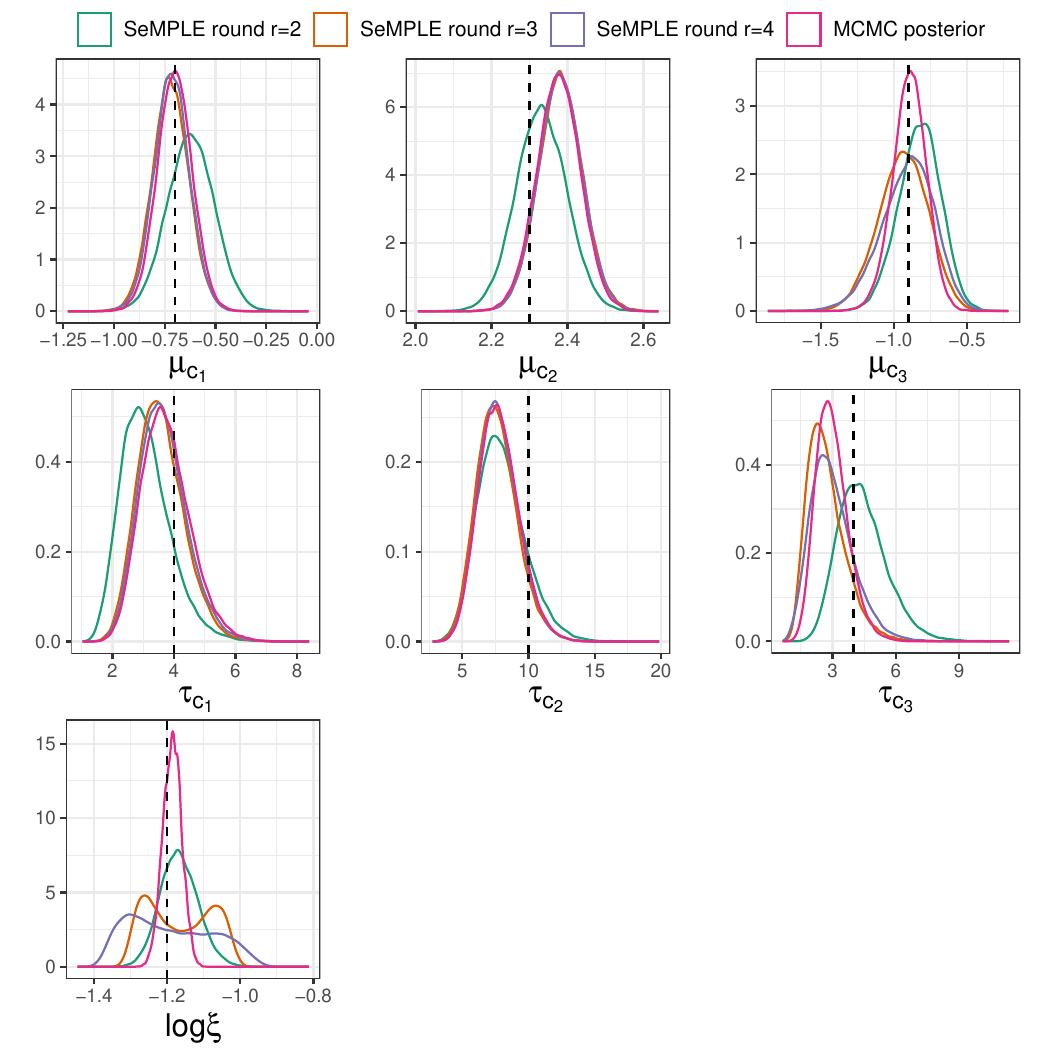}
    \caption{Ornstein-Uhlenbeck, inference from SeMPLE rounds $r=2,3,4$, and MCMC posteriors (when using the exact likelihood). The dashed lines show the true parameter values used to simulate the observed data.}
    \label{fig:OU_R4}
\end{figure}

\begin{figure}[H]
    \centering
    \includegraphics[width=0.8\linewidth]{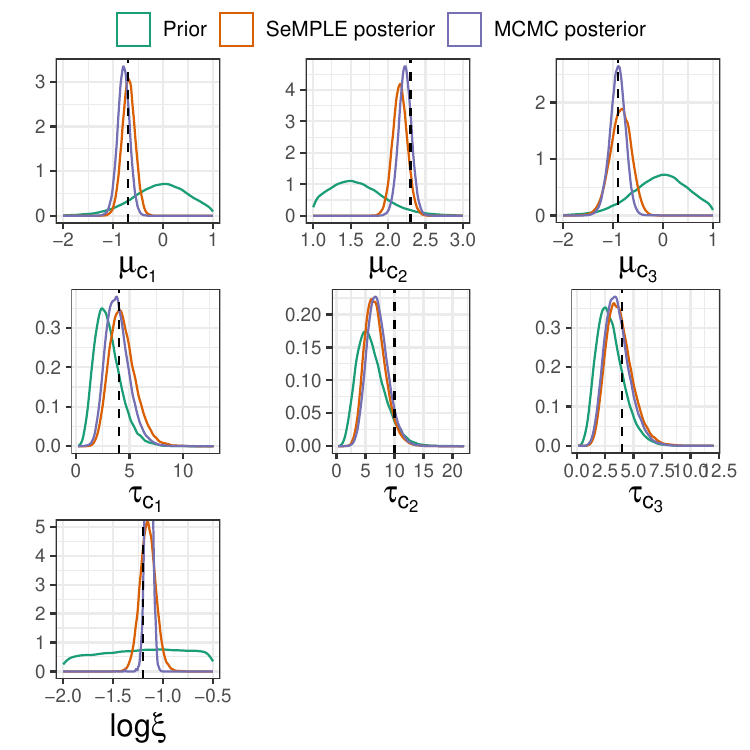}
    \caption{Ornstein-Uhlenbeck, inference with simulated data from round $r=4$ of SeMPLE when the number of individuals is $M=20$.}
    \label{fig:OU_M20}
\end{figure}
\begin{figure}[H]
    \centering
    \includegraphics[width=0.8\linewidth]{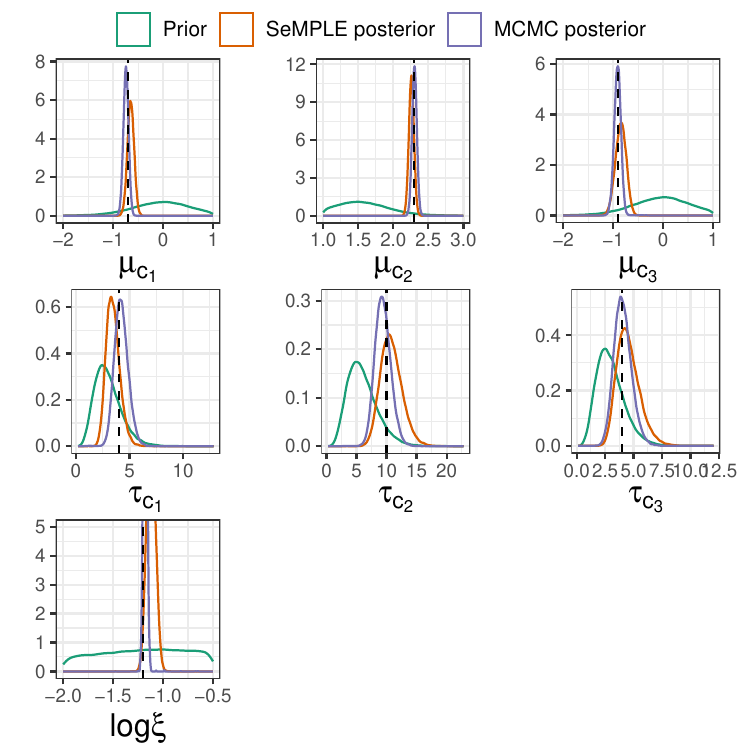}
    \caption{Ornstein-Uhlenbeck, inference with simulated data from round $r=4$ of SeMPLE when the number of individuals is $M=100$.}
    \label{fig:OU_M100}
\end{figure}

\subsection{mRNA model: simulated data} \label{sec:app_mrna_results}
The prior distributions are set to be Normal-Gamma for the random effects
\begin{align}
    \begin{cases}
    \mu_j | \tau_j \sim \mathcal{N}(\mu_{0_j}, (\lambda_j\tau_j)^{-1}) \label{eq:mrna_prior}\\ 
    \tau_j \sim Ga(\alpha_j, \beta_j),   \qquad j=\delta,\gamma,k,
    \end{cases}
\end{align}
where $\beta_j$ is a rate parameter and the population prior parameter values can be found in Table \ref{tab:mrna_prior_parameters}
\begin{table}[h]
\begin{tabular}{l|llll}
j & $\mu_{0_j}$ & $\lambda_j$ & $\alpha_j$ & $\beta_j$ \\
\hline
$\delta$  &     -0.694& 1& 2& 0.5 \\
$\gamma$  &    -3& 1& 2& 0.5 \\
$k$  &        0.027& 1& 2& 0.5
\end{tabular}
\caption{mRNA model simulated data: prior distribution parameters.}
\label{tab:mrna_prior_parameters}
\end{table}
and the fixed effects have Gaussian priors on the log-scale
    $\log m_0 \sim \mathcal{N}(5, 1)$,  
    $\log \mathrm{scale} \sim \mathcal{N}(1, 1)$,  
    $\log \mathrm{offset} \sim \mathcal{N}(3, 1)$,  
    $\log \xi \sim \mathcal{N}(-1.5, 1)$.

\begin{figure}[H]
    \centering
    \begin{subfigure}[b]{0.475\linewidth}
        \centering
        \includegraphics[width=0.7\linewidth]{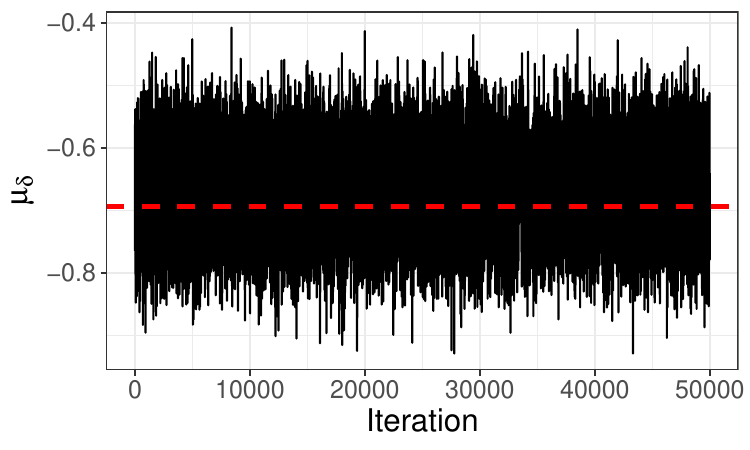}
        \caption{$\mu_\delta$}
    \end{subfigure}
        \begin{subfigure}[b]{0.475\linewidth}
        \centering
        \includegraphics[width=0.7\linewidth]{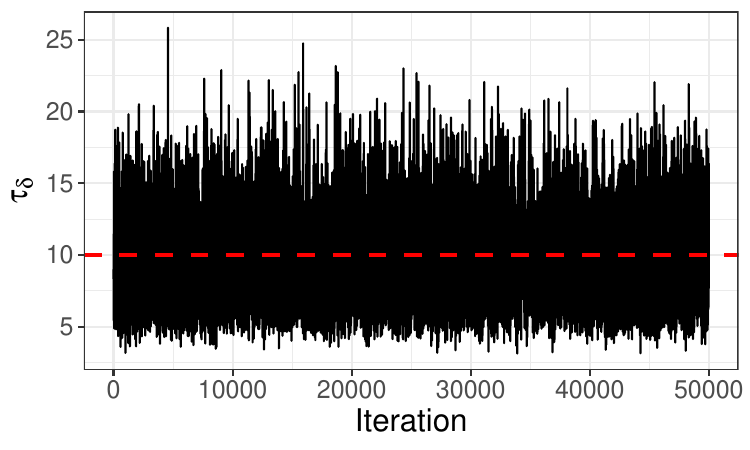}
        \caption{$\tau_\delta$}
    \end{subfigure}
    \begin{subfigure}[b]{0.475\linewidth}
        \centering
        \includegraphics[width=0.7\linewidth]{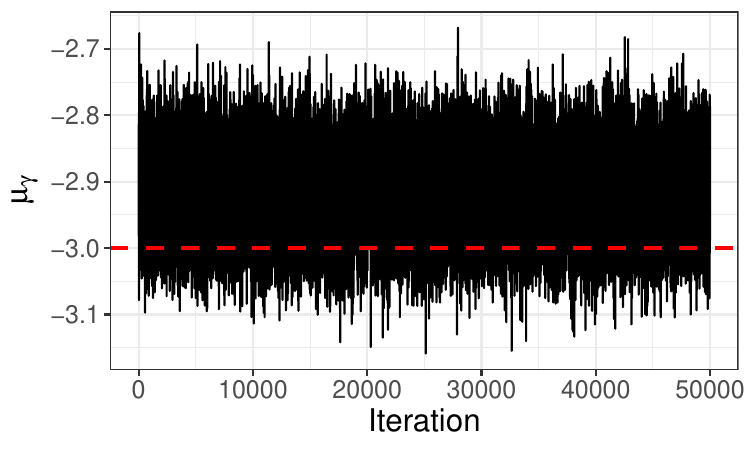}
        \caption{$\mu_\gamma$}
    \end{subfigure}
    \begin{subfigure}[b]{0.475\linewidth}
        \centering
        \includegraphics[width=0.7\linewidth]{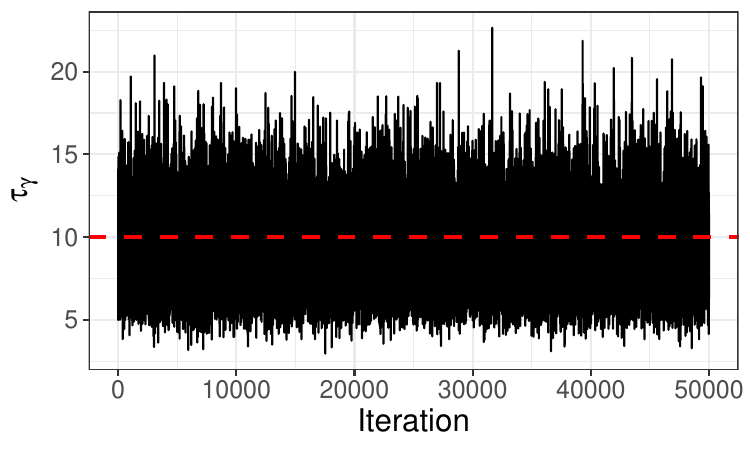}
        \caption{$\tau_\gamma$}
    \end{subfigure}
    \begin{subfigure}[b]{0.475\linewidth}
        \centering
        \includegraphics[width=0.7\linewidth]{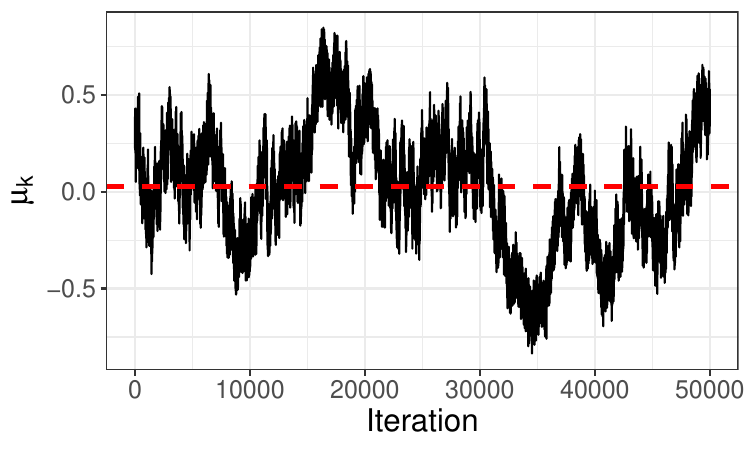}
        \caption{$\mu_k$}
    \end{subfigure}
    \begin{subfigure}[b]{0.475\linewidth}
        \centering
        \includegraphics[width=0.7\linewidth]{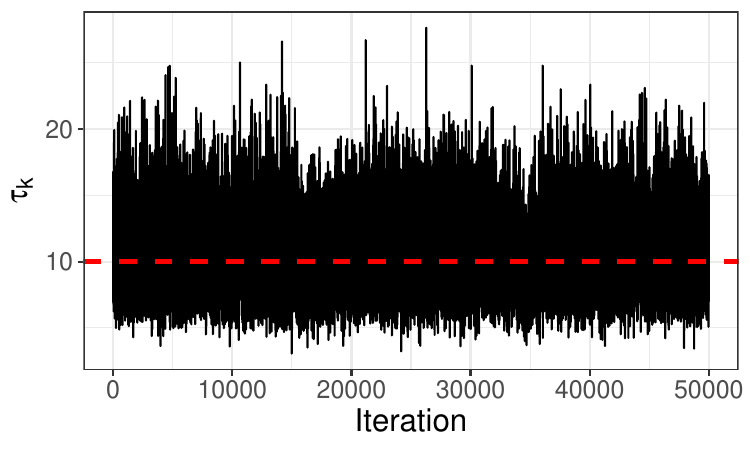}
        \caption{$\tau_k$}
    \end{subfigure}
    \begin{subfigure}[b]{0.475\linewidth}
        \centering
        \includegraphics[width=0.7\linewidth]{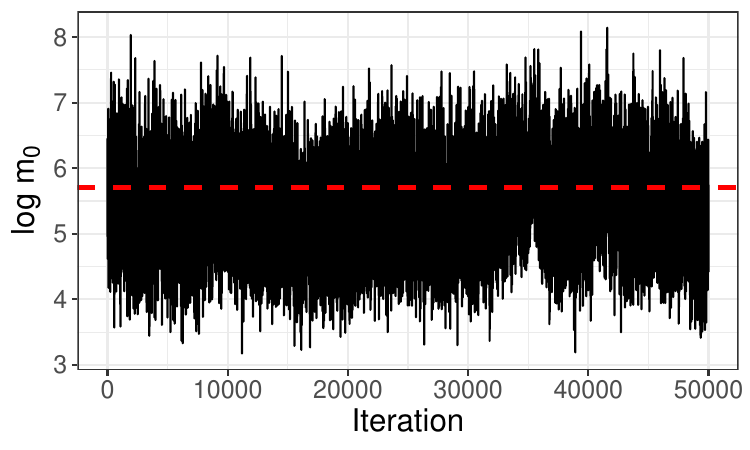}
        \caption{$\log m_0$}
    \end{subfigure}
        \begin{subfigure}[b]{0.475\linewidth}
        \centering
        \includegraphics[width=0.7\linewidth]{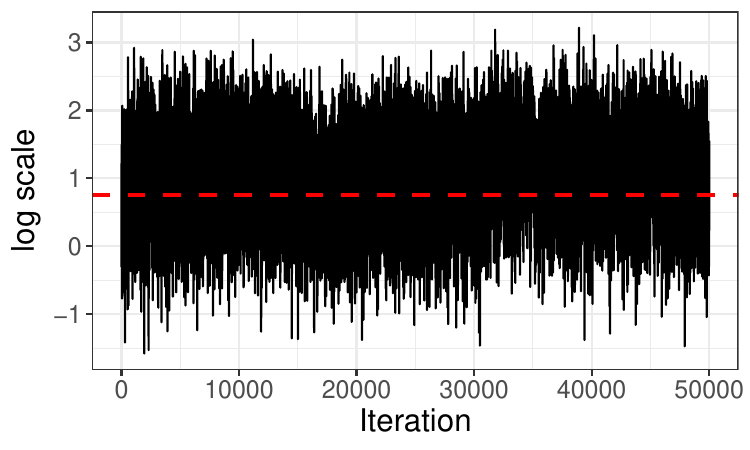}
        \caption{$\log \mathrm{scale}$}
    \end{subfigure}
    \begin{subfigure}[b]{0.475\linewidth}
        \centering
        \includegraphics[width=0.7\linewidth]{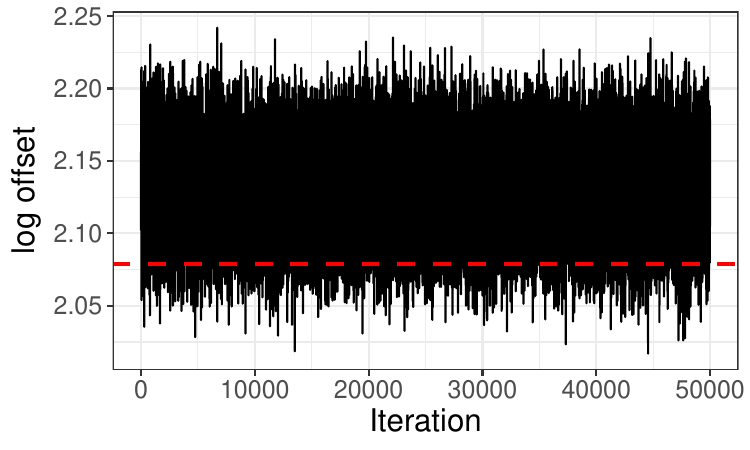}
        \caption{$\log \mathrm{offset}$}
    \end{subfigure}
    \begin{subfigure}[b]{0.475\linewidth}
        \centering
        \includegraphics[width=0.7\linewidth]{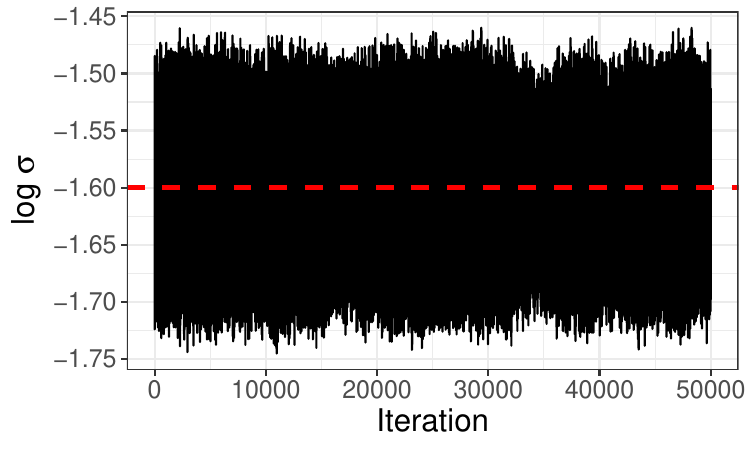}
        \caption{$\log \sigma$}
    \end{subfigure}
    \caption{mRNA model with simulated data: SeMPLE traceplots of population parameters $\bseta = (\mu_\delta, \mu_\gamma, \mu_k, \tau_\delta, \tau_\gamma, \tau_k)$, $\kappa = (m_0, \mathrm{scale}, \mathrm{offset})$ and $\bxi = \sigma$ from SeMPLE round $r=4$. The red dashed lines shows the true parameter values.}
    \label{fig:mrna_sim_traceplot_eta_kappaxi}
\end{figure}

\begin{figure}[H]
    \centering
    \begin{subfigure}[b]{0.49\linewidth}
        \centering
        \includegraphics[width=0.7\linewidth]{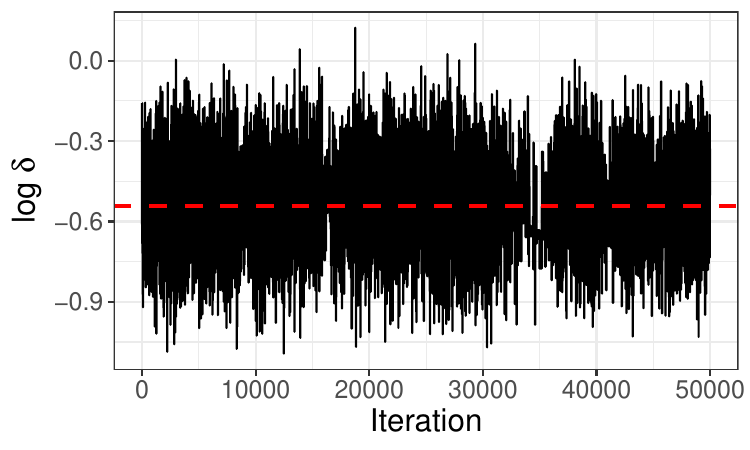}
        \caption{$\log \delta^{(1)}$}
    \end{subfigure}
        \begin{subfigure}[b]{0.49\linewidth}
        \centering
        \includegraphics[width=0.7\linewidth]{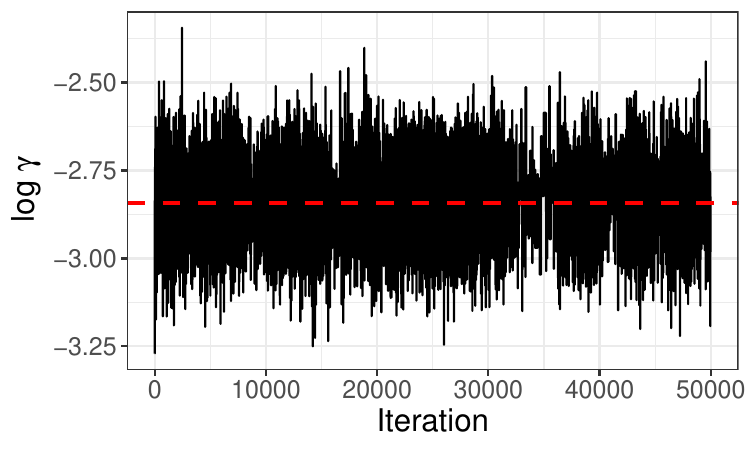}
        \caption{$\log \gamma^{(1)}$}
    \end{subfigure}
    \begin{subfigure}[b]{0.49\linewidth}
        \centering
        \includegraphics[width=0.7\linewidth]{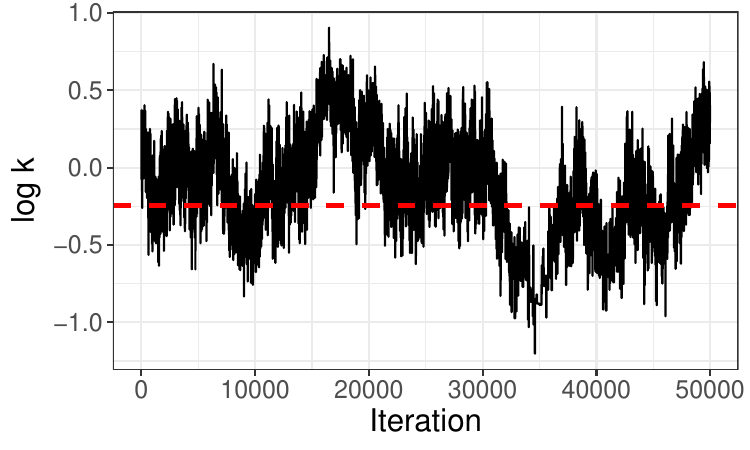}
        \caption{$\log k^{(1)}$}
    \end{subfigure}
    \caption{mRNA model with simulated data, traceplots for individual parameters from round $r=4$ of SeMPLE for individual 1 out of 40. The red dashed lines show the true parameters for this specific individual.
    }
    \label{fig:mrna_sim_traceplot_c}
\end{figure}

\begin{figure}[H]
    \centering
    \begin{subfigure}[b]{0.475\linewidth}
        \centering
        \includegraphics[width=0.7\linewidth]{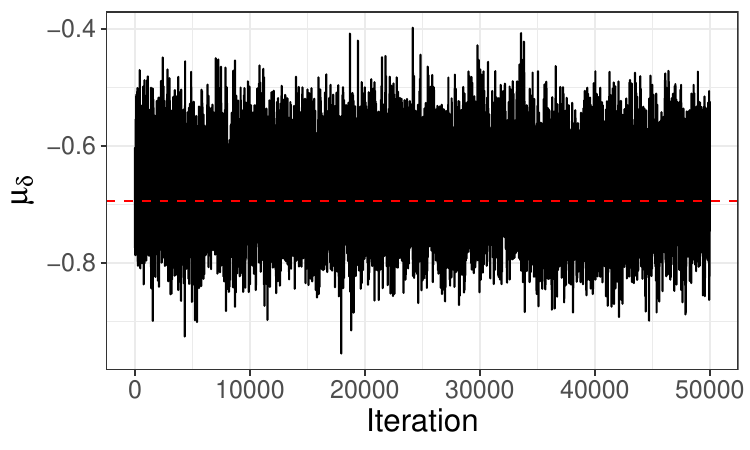}
        \caption{$\mu_\delta$}
    \end{subfigure}
        \begin{subfigure}[b]{0.475\linewidth}
        \centering
        \includegraphics[width=0.7\linewidth]{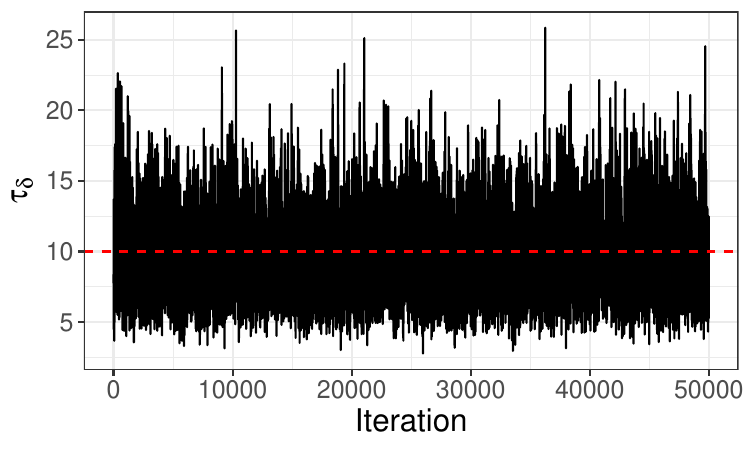}
        \caption{$\tau_\delta$}
    \end{subfigure}
    \begin{subfigure}[b]{0.475\linewidth}
        \centering
        \includegraphics[width=0.7\linewidth]{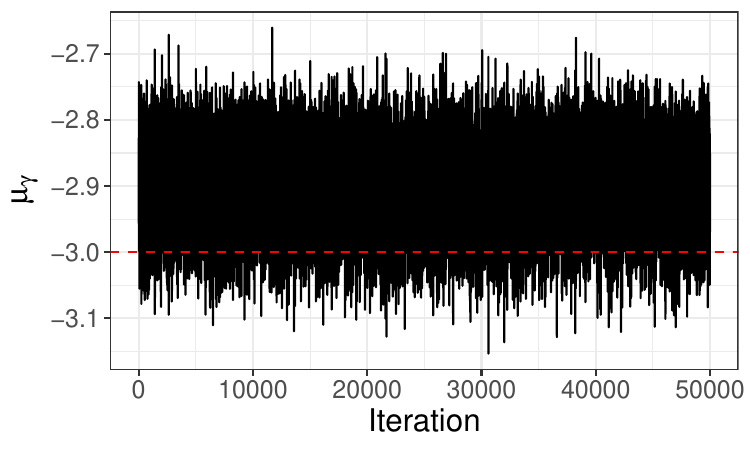}
        \caption{$\mu_\gamma$}
    \end{subfigure}
        \begin{subfigure}[b]{0.475\linewidth}
        \centering
        \includegraphics[width=0.7\linewidth]{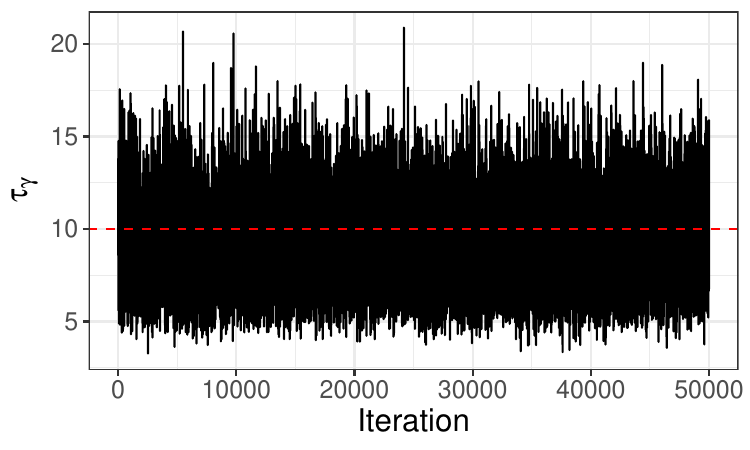}
        \caption{$\tau_\gamma$}
    \end{subfigure}
    \begin{subfigure}[b]{0.475\linewidth}
        \centering
        \includegraphics[width=0.7\linewidth]{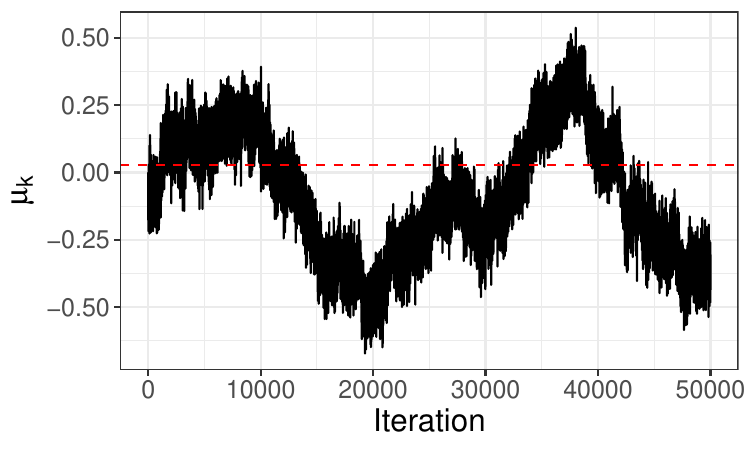}
        \caption{$\mu_k$}
    \end{subfigure}
        \begin{subfigure}[b]{0.475\linewidth}
        \centering
        \includegraphics[width=0.7\linewidth]{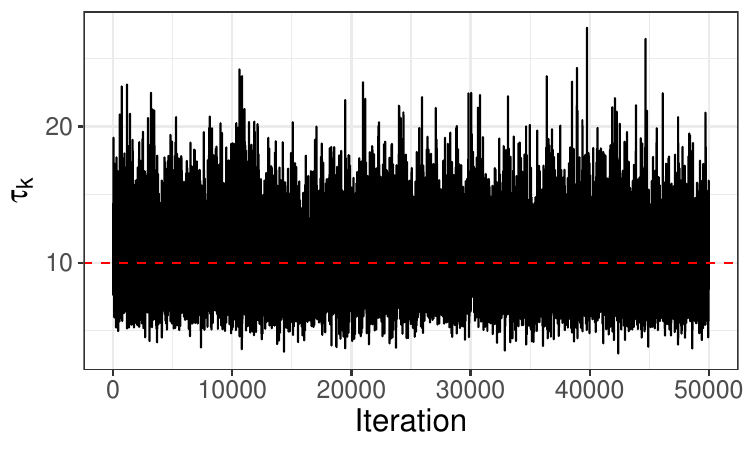}
        \caption{$\tau_k$}
    \end{subfigure}

    \begin{subfigure}[b]{0.475\linewidth}
        \centering
        \includegraphics[width=0.7\linewidth]{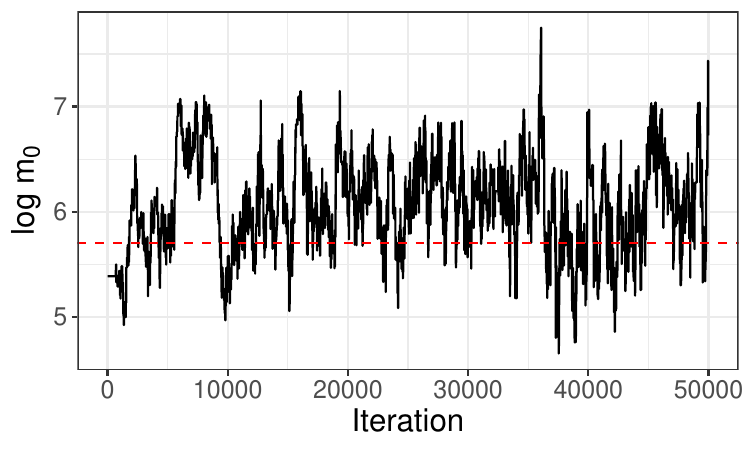}
        \caption{$\log m_0$}
    \end{subfigure}
        \begin{subfigure}[b]{0.475\linewidth}
        \centering
        \includegraphics[width=0.7\linewidth]{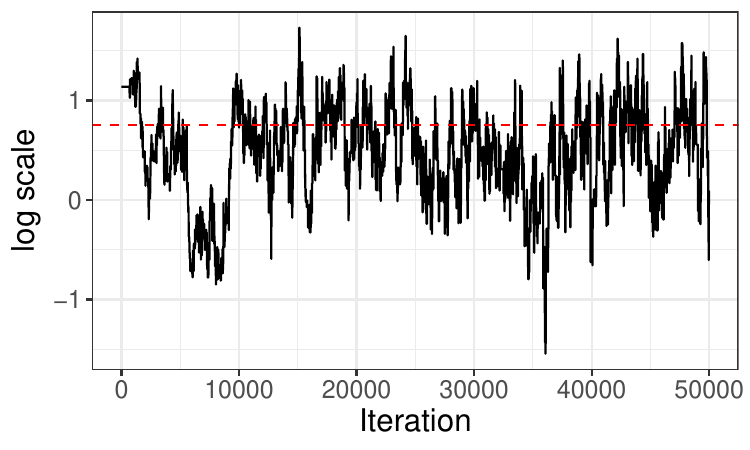}
        \caption{$\log \mathrm{scale}$}
    \end{subfigure}
    \begin{subfigure}[b]{0.475\linewidth}
        \centering
        \includegraphics[width=0.7\linewidth]{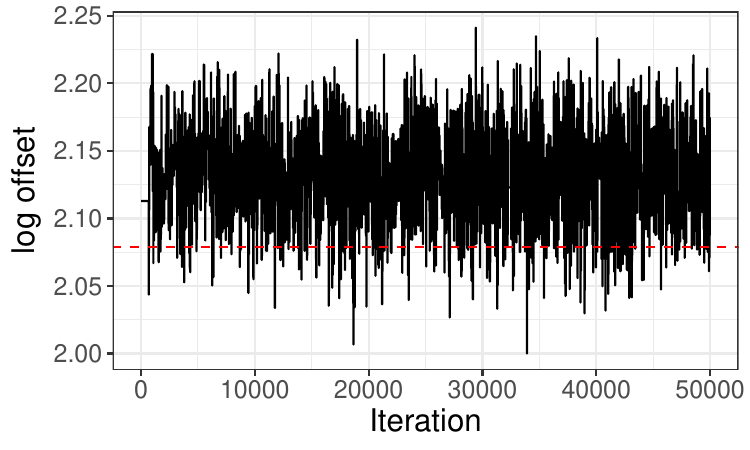}
        \caption{$\log \mathrm{offset}$}
    \end{subfigure}
    \begin{subfigure}[b]{0.475\linewidth}
        \centering
        \includegraphics[width=0.7\linewidth]{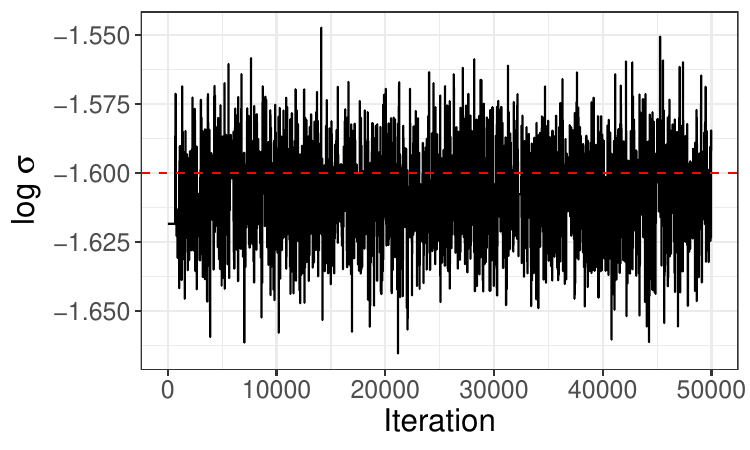}
        \caption{$\log \sigma$}
    \end{subfigure}
    \caption{mRNA model with simulated data: PEPSDI traceplots of parameters $\bseta = (\mu_\delta, \mu_\gamma, \mu_k, \tau_\delta, \tau_\gamma, \tau_k)$, $\kappa = (m_0, \mathrm{scale}, \mathrm{offset})$ and $\bxi = \sigma$. The red dashed line shows the true parameter values.}
    \label{fig:mrna_sim_pepsdi_traceplot_kappaxi}
\end{figure}

\begin{figure}[H]
    \centering
    \begin{subfigure}[b]{0.55\linewidth}
        \centering
        \includegraphics[width=\linewidth]{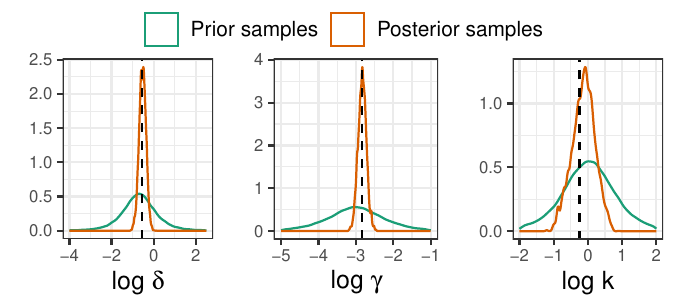}
    \end{subfigure}
    \begin{subfigure}[b]{0.55\linewidth}
        \centering
        \includegraphics[width=\linewidth]{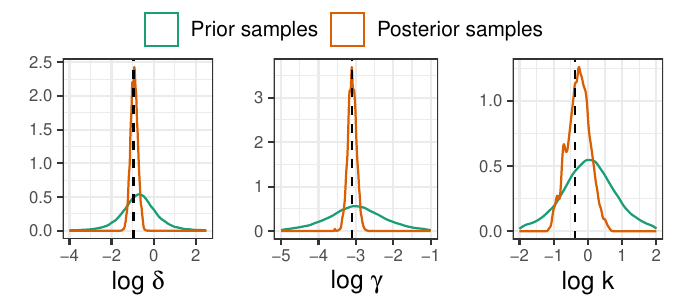}
    \end{subfigure}
    \begin{subfigure}[b]{0.55\linewidth}
        \centering
        \includegraphics[width=\linewidth]{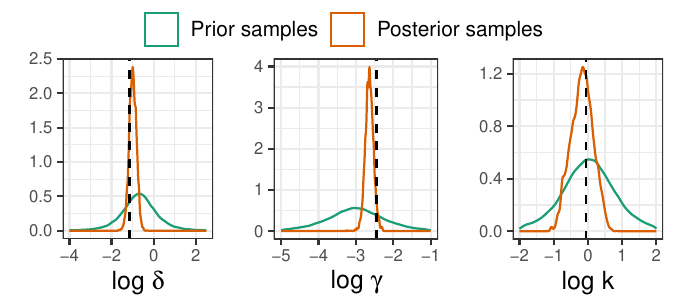}
    \end{subfigure}
    \begin{subfigure}[b]{0.55\linewidth}
        \centering
        \includegraphics[width=\linewidth]{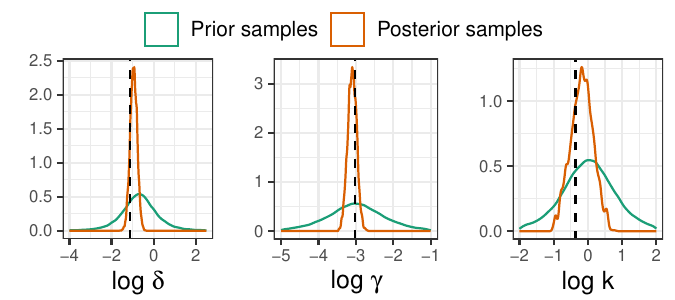}
    \end{subfigure}
    \begin{subfigure}[b]{0.55\linewidth}
        \centering
        \includegraphics[width=\linewidth]{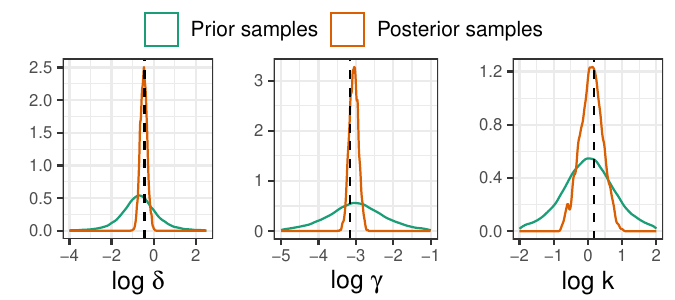}
    \end{subfigure}
    \label{fig:mrna_kde_individual}
    \caption{mRNA with simulated data: marginal posteriors from round $r=4$ of SeMPLE for the individual parameters $(\log\delta^{(i)},\log\gamma^{(i)},\log\kappa^{(i)})$ of the first 5 individuals out of 40. The black dashed lines shows the true parameter values of each specific individual.}
\end{figure}

\begin{figure}[H]
    \centering
    \includegraphics[width=0.9\linewidth]{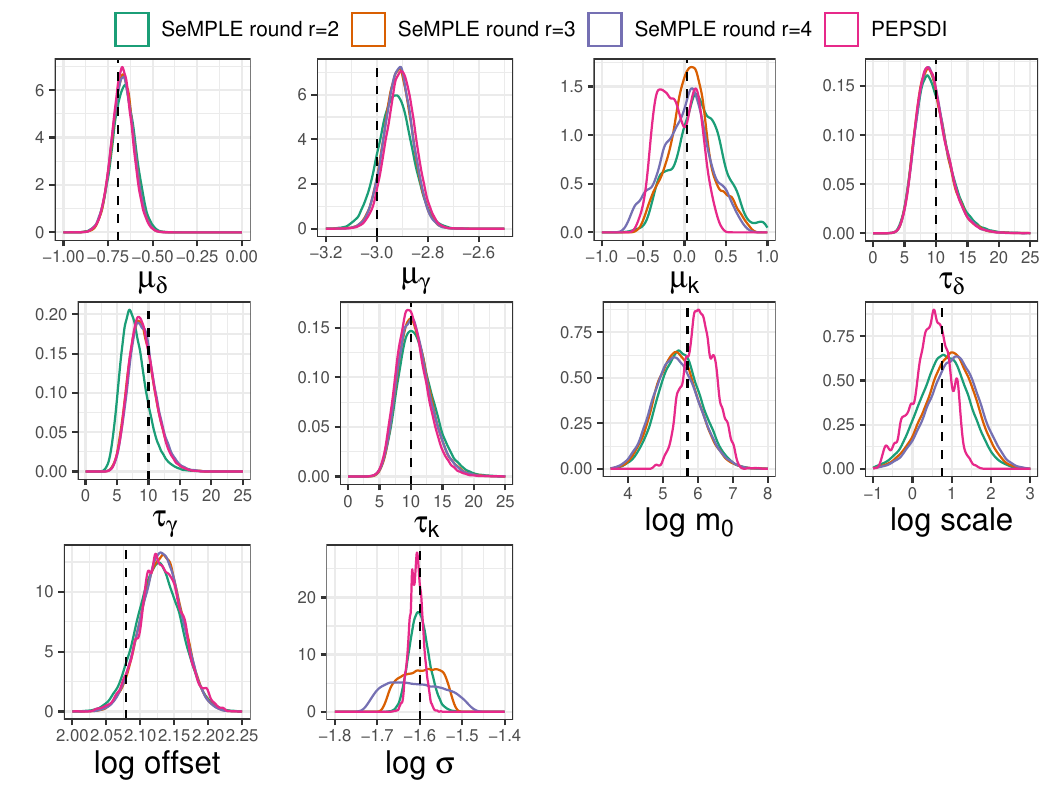}
    \caption{mRNA model with simulated data from $M=40$ individuals: marginal posteriors from round $r=2,3,4$ of SeMPLE and the PEPSDI reference samples.}
    \label{fig:mrna_kde_R4}
\end{figure}

\begin{figure}[H]
    \centering
    \includegraphics[width=0.5\linewidth]{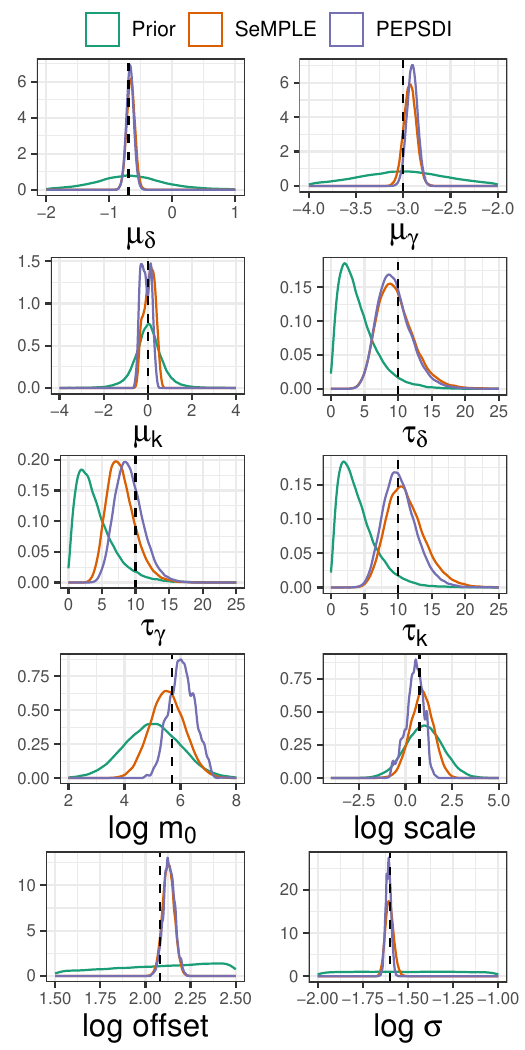}
    \caption{mRNA model with 40 simulated individuals: marginal posteriors obtained with SeMPLE (orange, round $r=2$) and with PEPSDI (purple). Priors are in green. The dashed lines are the true parameter values that were used to generate the observed data.}
    \label{fig:mrna_kde_comp_pepsdi_r2}
\end{figure}

\begin{table}[]
\scalebox{0.9}{
\begin{tabular}{llllllllllll}
                               & $\mu_\delta$ & $\mu_\gamma$ & $\mu_k$ & $\tau_\delta$ & $\tau_\gamma$ & $\tau_k$ & $\log m_0$ & $\log$ offset & $\log$ scale & $\log \sigma$ &  \\ \hline

\textbf{SeMPLE at round 4}                &&&&& &&&&& &     \\
ESS                            & 8010 & 18100 & 17.8 & 7130 & 11600 & 6900 & 2170 & 3510 & 24000 & 8700 &  \\
ESS/sec                        & 0.0288 & 0.0649 & 6.39e-05 & 0.0256 & 0.0415 & 0.0248 & 0.00779 & 0.0126 & 0.0864 & 0.0312 &       \\
ESS multivariate               &&&&& &&&&& & 7590 \\

                               & & & & & & & & & & & \\ \hline

\textbf{PEPSDI}                &              &              &         &               &               &          &            &        &       &          \\
ESS                            &   4400 & 9250 & 5.37 & 2520 & 8720 & 4510 & 83.4 & 75 & 1210 & 1680        \\
ESS/sec                        &      0.0082 & 0.0173 & 1e-05 & 0.00469 & 0.0163 & 0.0084 & 0.000156 & 0.00014 & 0.00225 & 0.00313        \\

ESS multivariate               &&&&& &&&&& & 1810\\
\end{tabular}
}
\caption{mRNA model with simulated data from $M=40$ individuals: Effective sample size (ESS), both univariate and multivariate, and ESS/sec, for SeMPLE (round $r=4$) and PEPSDI.}
\label{tab:ess_iat}
\end{table}

\begin{figure}[H]
    \centering
    \includegraphics[width=\linewidth]{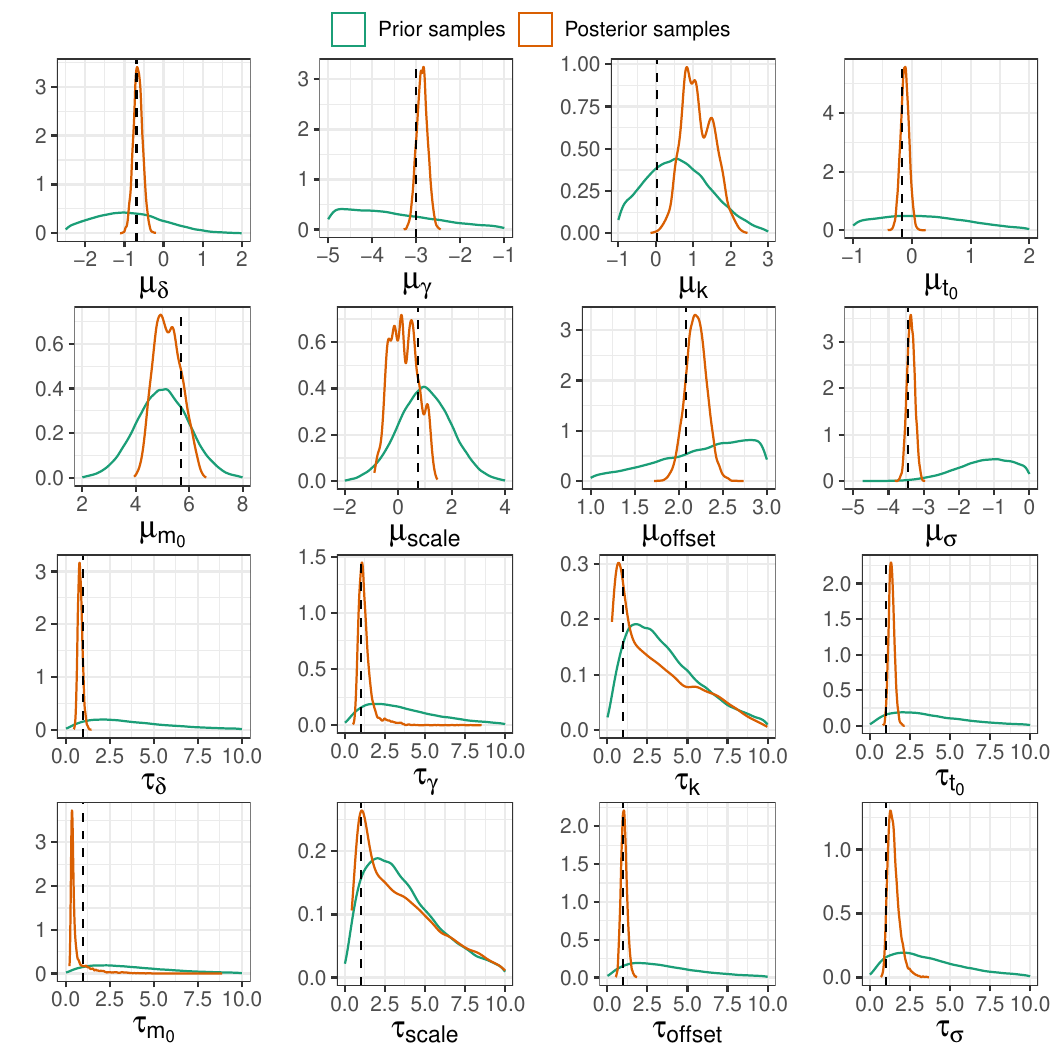}
    \caption{mRNA model with simulated data from $M=200$ individuals when assuming that all parameters are random effects: marginal posteriors obtained with SeMPLE (orange, round $r=4$) and priors (green). Note that the true population precision $\btau = 1$ for all parameters. The marginal posteriors are based on 10k posterior samples.}
    \label{fig:mrna_kde_200ind}
\end{figure}

\section{mRNA model: real data} \label{sec:app_mrna_real_data}
In the inference setup with real data the priors are set to
\begin{align}
    \begin{cases}
    \mu_j \sim \mathcal{N}(\mu_{0_j}, \sigma_j^2) \label{eq:mrna_prior_indep}\\ 
    \tau_j \sim Ga(\alpha_j, \beta_j),   \qquad j=\delta,\gamma,k,t_0.
    \end{cases}
\end{align}
where the population prior parameters can be found in Table \ref{tab:mrna_real_data_prior_parameters} and $\log m_0 \sim \mathcal{N}(5, 1)$, 
     $\log \mathrm{scale} \sim \mathcal{N}(1, 1)$, 
     $\log \mathrm{offset} \sim \mathcal{N}(3, 1)$, 
     $\log \sigma \sim \mathcal{N}(-1, 1)$.
\begin{table}[h]
\begin{tabular}{l|llll}
j & $\mu_{0_j}$ & $\sigma_j$ & $\alpha_j$ & $\beta_j$ \\
\hline
$\delta$  &     -1& 1& 2& 0.5 \\
$\gamma$  &    -5& 2& 2& 0.5 \\
$k$  &        0.5& 1& 2& 0.5 \\
$t_0$ &       0& 1& 2& 0.5
\end{tabular}
\caption{mRNA model real data: population prior distribution parameters.}
\label{tab:mrna_real_data_prior_parameters}
\end{table}

\begin{figure}[H]
    \centering
    \begin{subfigure}[b]{0.49\linewidth}
        \centering
        \includegraphics[width=0.7\linewidth]{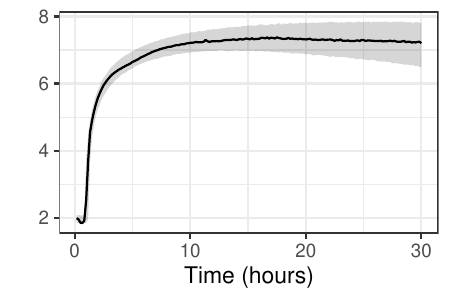}
    \end{subfigure}
    \begin{subfigure}[b]{0.49\linewidth}
        \centering
        \includegraphics[width=0.7\linewidth]{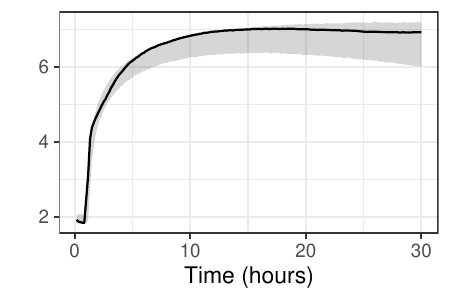}
    \end{subfigure}
    \begin{subfigure}[b]{0.49\linewidth}
        \centering
        \includegraphics[width=0.7\linewidth]{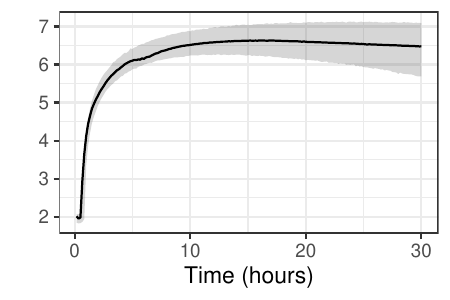}
    \end{subfigure}
    \begin{subfigure}[b]{0.49\linewidth}
        \centering
        \includegraphics[width=0.7\linewidth]{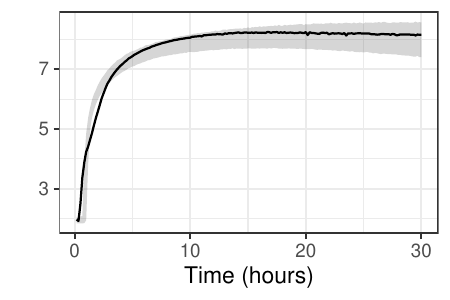}
    \end{subfigure}
    \begin{subfigure}[b]{0.49\linewidth}
        \centering
        \includegraphics[width=0.7\linewidth]{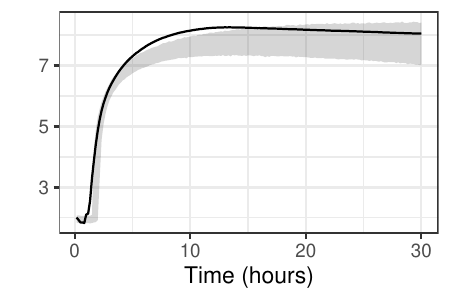}
    \end{subfigure}
    \caption{mRNA model with real data: Posterior-predictive simulations based on 1k individual specific SeMPLE posterior samples for the first 5 individuals (top left to bottom rigth). The black line shows the observed trajectory of each individual. In gray is the area between the 2.5th and 97.5th percentile.}
    \label{fig:mrna_ppc_individual}
\end{figure}

\begin{figure}
    \centering
    \includegraphics[width=0.95\linewidth]{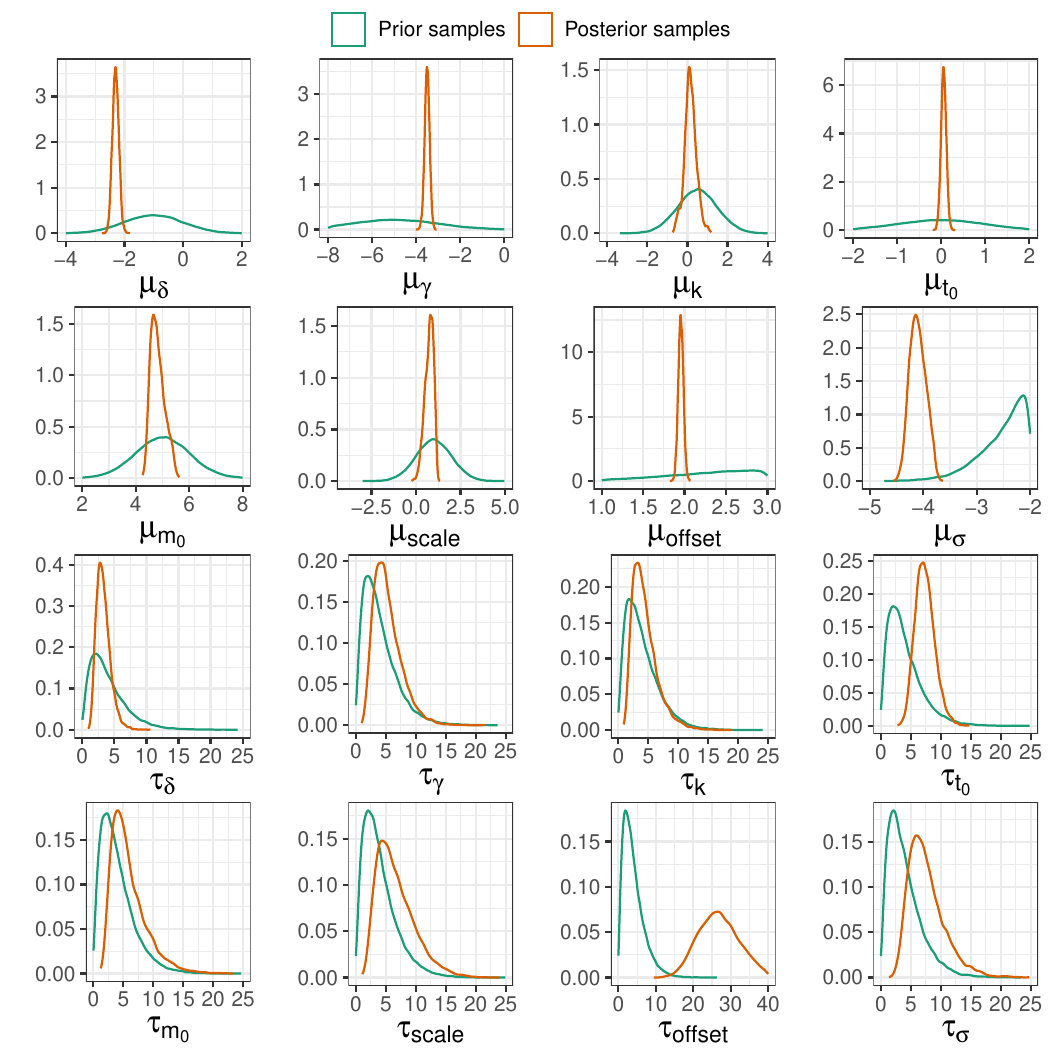}
    \caption{mRNA model with real data when assuming that all parameters are random effects: marginal posteriors obtained with SeMPLE (orange, round $r=4$) and priors (green).}
    \label{fig:mrna_kde_only_individual}
\end{figure}

\end{document}